\documentclass[11pt,a4paper]{article}

\usepackage{jheppub}
\usepackage{amsmath,amsfonts,amsthm,bm}

\usepackage{color}

\def\nn{\nonumber}
\def\o{{\cal O}}
\def\R{{\cal R}}
\def\l{\lambda}
\def\r{\rho}
\def\g{\gamma}
\def\rtt{\rho_{tt}}
\def\rtc{\rho_{tc}}
\def\rct{\rho_{ct}}
\def\cg{c_\gamma}

\def\eps{\varepsilon_K}
\def\epsp{\varepsilon^\prime/\varepsilon}
\def\KLnn{K_L \to \pi^0\nu\bar\nu}
\def\Kpnn{K^+ \to \pi^+\nu\bar\nu}
\def\KLmm{K_L \to \mu^+\mu^-}
\def\KSmm{{K_S \to \mu^+\mu^-}}

\title{
Strange processes in general two Higgs doublet model
}
\author{Wei-Shu Hou and Girish Kumar}
\affiliation{
Department of Physics, National Taiwan University, Taipei 10617, Taiwan
}
\emailAdd{wshou@phys.ntu.edu.tw}
\emailAdd{girishk@hep1.phys.ntu.edu.tw}

\abstract{
In the general two Higgs doublet model (g2HDM) that has 
extra Yukawa couplings, we analyze their New Physics (NP)
contributions to kaon mixing parameter $\eps$, 
direct CP violation parameter $\epsp$, and rare $\Kpnn$, 
$\KLnn$, and $K_{L, S} \to \mu^+ \mu^-$ decays. We study 
correlations between these observables, and call special 
attention to a unique complementarity of kaon mixing and 
rare $K\to \pi \nu\bar\nu$ decays in probing the exotic 
Higgs mass spectrum of g2HDM. The importance of kaon physics 
in probing NP vis-\`a-vis B physics is stressed.
One unexpected feature we uncover is the special 
sensitivity of $\Kpnn$ to TeV scale charged Higgs boson.}

\arxivnumber{2207.07030}
\begin{document}
\maketitle
\flushbottom

\section{Introduction}\label{sec: intro}

The general two Higgs doublet model (g2HDM) is one of the 
simplest NP models where one augments the Standard Model 
(SM) with another scalar SU(2) doublet~\cite{Lee:1973iz} 
(see Ref.~\cite{Branco:2011iw} for a review of 2HDMs). In 
contrast to the typical Type II 2HDM, which adopts an 
\emph{ad hoc} $Z_2$ symmetry on Yukawa interactions to 
enforce the natural flavor conservation 
condition~\cite{Glashow:1976nt}, no such discrete 
symmetry is imposed in g2HDM \cite{Hou:1991un}.  Consequently, in g2HDM there exists {\it extra} Yukawa 
couplings which are generic in size and complex in 
nature, with flavor-changing neutral Higgs (FCNH) 
couplings controlled by fermion mass-mixing hierarchies 
{that are} built-in by {\it Nature} herself. Of 
particular interest are extra Yukawa couplings related 
to the top quark, $\rho_{tt}$ and $\rho_{tc}$, where 
they can provide new sources for charge-parity violation 
(CPV), which together with ${\o}(1)$ Higgs quartic 
couplings~\cite{Kanemura:2004ch}, can satisfy the 
Sakharov conditions~\cite{Sakharov:1967dj} and give rise 
to electroweak baryogenesis (EWBG), i.e. account for the 
baryon asymmetry of the Universe (BAU), as illustrated 
in Refs.~\cite{Fuyuto:2017ewj, Fuyuto:2019svr}.

An appealing aspect of EWBG in g2HDM is its testability 
at ongoing and upcoming collider and flavor experiments. 
For example, the couplings $\rho_{tt}$ and $\rho_{tc}$ 
at ${\cal O}(\lambda_t)$ strength with $\l_t \cong 1$ 
the SM top quark Yukawa coupling, can propel exquisite 
collider signatures such as~\cite{Kohda:2017fkn} $cg \to 
t H/tA \to tt\bar c$ (same-sign top plus c-jet) and 
$tt\bar t$ (triple-top), where $H/A$ are CP-even/odd 
exotic neutral scalar bosons in g2HDM. Another important 
probe is charged Higgs ($H^+$) associated production: 
$cg \to bH^+ \to bt\bar b$ \cite{Ghosh:2019exx}. Besides 
$\rho_{tc}$ and $\rho_{tt}$ at ${\cal O}(1)$, this 
process is further enhanced at the amplitude level --- 
in contrast to Type II 2HDM --- by a CKM ratio 
$V_{tb}/V_{cb}\sim 24$,
making it a unique probe of g2HDM. A review on prospects 
at LHC and the High-Luminosity LHC (HL-LHC) can be found 
in Ref.~\cite{Hou:2020chc}.

Another most promising probe of $\rho_{tt}$ in the context 
of EWBG is the electric dipole moment (EDM) of the 
electron. An upper limit of $|d_e| < 1.1 \times 10^{-29} 
\;e\,{\rm cm}$ is set~\cite{ACME:2013pal, ACME:2018yjb} 
by ACME, which is the best limit on any EDM. The coupling 
$\rho_{tt}$ contributes to electron EDM via two-loop 
(Barr-Zee) diagrams~\cite{Barr:1990vd}. A hierarchy --- 
similar to the one present in the SM --- between g2HDM 
couplings of top and electron $|\r_{ee}/\rtt| \propto 
\l_e/\l_t$ helps one evade~\cite{Fuyuto:2019svr} the 
ACME bound. Such direct correlations link EWBG realized 
at the very early Universe, to electron EDM being measured 
currently in the laboratories.

Similarly, quark flavor observables provide important
probes of $\rho_{ij}$ couplings. In the literature, 
B physics observables in particular have been 
discussed frequently to constrain g2HDM couplings.
For example, precise measurements of neutral $B_q$ ($q=s, d$) mixings \cite{HFLAV:2019otj} put stringent
constraints on $\rho_{ij}$ \cite{Altunkaynak:2015twa};
in particular, the off-diagonal down-type couplings 
get severely constrained \cite{Crivellin:2013wna}.
The inclusive radiative decay $B\to X_s \gamma $ 
is known to be one of the most sensitive probes of
$H^+$. The measured value of its branching ratio
${\cal B}(B\to X_s \gamma)=(3.32 \pm 0.15) \times
10^{-4}$ \cite{HFLAV:2019otj}
%
%
agrees well with its SM prediction \cite{Misiak:2015xwa}.
For Type II 2HDM, the decay already sets the limit 
$m_{H^+} > 580$~GeV at $95\%$ C.L. \cite{Misiak:2017bgg}. 
In contrast, $b \to s \gamma$ easily accommodates lower
values of $H^+$ in g2HDM, but seriously constrains the 
parameter space for $\rho_{ct}$ and the down coupling $\rho_{bb}$~\cite{Altunkaynak:2015twa,Hou:2022b2sgamma}. 

Another often discussed process in the literature is the 
rare $B_s \to \mu^+\mu^-$ decay. It is helicity-suppressed 
in the SM and therefore provides one of the most sensitive 
probes of scalar interactions. LHCb reported ${\cal B}(B_s 
\to \mu^+\mu^-) = (3.09^{+0.46\,+ 0.15}_{-0.43\, -0.11}) 
\times 10^{-9}$~\cite{LHCb:2021vsc, LHCb:2021awg}, based 
on full dataset collected during Run 1 and Run 2 with 
integrated luminosity of 9~${\rm fb}^{-1}$ in total. On 
the other hand, CMS has just reported their full Run 2 
analysis based on 2016-2018 data, corresponding to an 
integrated luminosity of 140~${\rm fb}^{-1}$, giving 
${\cal B}(B_s\to \mu^+\mu^-) = (3.83^{+0.38 \,+ 0.19 \,+ 
0.14}_{-0.36\,-0.16\, -0.13}) \times 10^{-9}$~\cite{CMS: 
ICHEP2022}, with central value about $1.2\sigma$ higher 
than LHCb result. These measurements agree with SM 
expectation~\cite{Bobeth:2013uxa, Beneke:2019slt} and 
{would} put strong constraints on up-type $\rho_{ij}$ 
couplings~\cite{Crivellin:2019dun, Iguro:2017ysu, 
Hou:2020itz}. A previous combined analysis of 
ATLAS~\cite{ATLAS:2018cur}, CMS~\cite{CMS:2019bbr} and 
LHCb~\cite{LHCb:2017rmj} based on 2011-2016 data found  ${\cal B}(B_s\to \mu^+\mu^-)_{\rm ave} = 
(2.69^{+0.37}_{-0.35})\times 10^{-9}$~\cite{LHCb:2020zud}, 
which is on the lower side of SM value, {but now the 
trend has changed. In our numerical analysis, we will 
take the LHCb result as reference value, but also discuss 
briefly the implication of the CMS update.} 

Interestingly enough, $b \to s\ell^+\ell^-$ data also exhibit 
significant tensions with SM in related $B \to K \ell\ell,\, 
K^\ast \ell \ell$ observables. Similarly, data related to 
charged current $b \to c \ell \nu$ also show deviations from SM. 
We refer to Ref.~\cite{London:2021lfn} for a recent summary 
about the \emph{B anomalies} (see Ref.~\cite{Hou:2019dgh} for 
an experimental critique). The g2HDM is capable of addressing 
several of these {B anomalies}, where again the top couplings 
$\rho_{tt}$, $\rho_{tc}$, and $\rho_{ct}$ play important roles. 
The $b \to s \ell\ell$ processes have been discussed in 
Refs.~\cite{Iguro:2018qzf, Crivellin:2019dun, Athron:2021auq} 
while $b \to c \ell \nu$ are discussed in 
Refs.~\cite{Iguro:2017ysu, Iguro:2022uzz, Blanke:2022pjy}. We 
caution that none of the deviations are confirmed individually. 
Future data may yet decide the fate of the B anomalies.

Despite the intense scrutiny of g2HDM interactions as briefly
outlined above, data still allows for sizable top-related
$\rho_{ij}$~\cite{Hou:2020chc}. We note, however, that a more
robust probe of flavor structure of any NP model is through
investigating correlations between NP contributions of 
different flavor sectors. This is particularly salient in 
case of g2HDM, where top-related $\rho_{ij}$ couplings that
contribute to flavor changing neutral coupling (FCNC) B 
processes would also affect other flavor sectors, most 
notably the kaon sector, which offers several observables 
that are very sensitive to NP contributions. We therefore
analyze g2HDM contributions to various kaon processes and investigate the prospects.

The study of kaon physics has been instrumental historically 
in shaping our current understanding of SM (see
Ref.~\cite{Buras:1998raa,Cirigliano:2011ny} for a review 
of kaon physics in SM). The $\varepsilon_K$ parameter of 
neutral kaon mixing is very precisely measured, and along
with mass differences $\Delta M_{B_q}$ $(q=s, d)$ are among
the most sensitive flavor probes of NP. The direct CPV
parameter $\epsp$ from $K \to \pi\pi$ decay is also a very
sensitive probe of CP violating NP~\cite{Buras:2014maa}. 
Back in 2015, RBC and UKQCD presented~\cite{RBC:2015gro} 
their first lattice QCD result for $K \to \pi\pi$ matrix 
elements and found $\epsp$ to be 2--$3\sigma$ below the 
experimental world average~\cite{NA48:2002tmj, KTeV:2002qqy, 
KTeV:2010sng}. This result received much attention and prompted 
several NP analyses, such as in g2HDM~\cite{Chen:2018ytc, 
Chen:2018vog, Iguro:2019zlc}. However, the 2020 
RBC-UKQCD~\cite{RBC:2020kdj} update of $A_0$ (isospin $I=0$) 
matrix element for $K \to \pi\pi$ gave $\epsp$ that is consistent 
with experiment. The theory uncertainties are still quite large, 
and significant NP contributions at {$\lesssim \o(10^{-3})$} may 
still be accommodated~\cite{Aebischer:2020jto}.

On the other hand, rare $\Kpnn$ and $\KLnn$ decays are
theoretically \emph{clean} processes, with branching 
ratios very precisely determined in SM. Dominated by 
$Z$ penguin and box diagrams in SM, they are highly
suppressed, making them very sensitive probes of NP 
scale --- even for scales beyond LHC reach. In contrast, 
rare decays $K_{L,S} \to \mu^+\mu^-$ have been less
enthusiastically pursued in the literature, as they 
receive both short-distance (SD) and long-distance (LD)
contributions; the LD contributions are dominated by two
on-shell photons~\cite{Ecker:1991ru,Isidori:2003ts,DAmbrosio:2017klp,Mescia:2006jd} and are quite significant. 
However, there has been important theoretical progress~\cite{DAmbrosio:2017klp,Dery:2021mct} 
recently towards a reliable extraction of SD parameters 
using data, making these processes, in particular $\KSmm$,
good probes of NP.

This paper is organized as follows. In the next section we 
introduce the Yukawa interactions in g2HDM and discuss the NP 
parameters relevant for our study. In Sec.~\ref{sec: kaon}, we 
discuss each of the aforementioned kaon observables in the 
context of g2HDM. Then, in Sec.~\ref{sec: results} we discuss 
constraints from B physics and present our results for kaon 
observables. We summarize our conclusions in Sec.~\ref{sec: 
summary}.

\section{Relevant Interactions in the Model}

As mentioned in the Introduction, g2HDM does not possess 
any discrete symmetry. Thus, 
%
the scalar doublets $\Phi_1$ and $\Phi_2$ are
indistinguishable, and both couple to up-type as well as
down-type SM fermions. Here, for convenience, we choose 
to work in the so-called Higgs basis, in which only one
doublet receives vacuum expectation value (vev): $\langle
\Phi_1\rangle \ne 0$, $\langle \Phi_2\rangle=0$. After
spontaneous symmetry breaking, the receiver of vev is
identified with the SM Higgs doublet and participates in
generating particle masses. The other doublet gives rise 
to new interactions with fermions {and interactions
between the two Higgs doublets. In} the physical basis, 
the Yukawa Lagrangian of the model is given 
as~\cite{Davidson:2005cw, Hou:2017hiw},
\begin{align}
\mathcal{L} = 
 -  \frac{1}{\sqrt{2}} \sum_{f = u, d, \ell} \bar f_{i}
    \Big[\big(\lambda^f_i \delta_{ij} s_\gamma 
            + \rho^f_{ij} c_\gamma\big) h 
      + \big(\lambda^f_i \delta_{ij} c_\gamma
            - \rho^f_{ij} s_\gamma\big) H
      - i\,{\rm sgn}(Q_f) \rho^f_{ij} A\Big] R\, f_{j} \nn\\
 - \bar{u}_i\left[(V\rho^d)_{ij}R-(\rho^{u\dagger}V)_{ij}L\right]d_j H^+
 - \bar{\nu}_i\rho^\ell_{ij} R \, \ell_j H^+
 +{h.c.},
\label{eq: Lag}
\end{align}
where we identify $h(125)$ with the discovered scalar at 
LHC, and $H$, $A$ and $H^\pm$ are exotic scalars which we
assume to be heavier than $h$. The couplings $\lambda_i =
\sqrt{2} m_i/v$ denote the SM Yukawa coupling, with $m_i$ 
the fermion mass and $v=246$ GeV the vev; $\rho_{ij}$ are
generic NP couplings, introduced already as extra Yukawa 
couplings; $R,\,L = (1 \pm \gamma_5)/2$ are chiral 
projections, and $V$ is the Cabibbo-Kobayashi-Maskawa (CKM)
matrix. The angle $\cg\equiv \cos \gamma$ ($s_\gamma \equiv
\sin\gamma$) describes the mixing between $h$ and $H$. 
The limit $\cg \to 0$, the so-called \emph{alignment limit},
provides an additional mechanism~\cite{Hou:2017hiw} for
suppressing FCNC involving the $h$ boson, which implies 
that the discovered $h$ boson approaches the SM Higgs boson.

As stated already, $\rho_{ij}$ are generic in size and contain
complex phases. However, the current data puts significant
constraints on their strength. For example, neutral meson 
mixing data severely constrains off-diagonal entries of down
type $\rho_{ij}^d$ couplings~\cite{Crivellin:2013wna},
indicating that $\rho^d$ matrix is almost diagonal in {\it 
Nature}. Further, data from proton-proton colliders such as 
the LHC already hint that $u$ and $d$-quark related couplings 
have to be small. Interestingly, and as stressed already,
the top-related couplings of g2HDM can be significantly 
larger~\cite{Hou:2020chc}. On the other hand, lepton related
$\rho_{ij}$ are less constrained. But given the assumption 
that $\rho_{tt}\sim {\cal O}(\lambda_t)$ and $\cg \sim {\cal
O}(0.1)$, data from flavor violating processes such as 
$h \to \tau \mu$, $\tau \to \mu \gamma$, $\mu \to e\gamma$ 
and $\mu\to e$ conversion in nuclei 
suggest~\cite{Hou:2020itz} $\rho_{\tau\tau},
\rho_{\tau\mu}\lesssim {\cal O}(\lambda_\tau)$ and
$\rho_{e\ell}\lesssim {\cal O}(\lambda_e)\, (\ell =e,
\mu,\tau)$, which are indeed very small.

We will therefore focus on the NP effects of only top- 
and charm-related couplings $\rho_{tt}$, $\rho_{tc}$, and
$\rho_{ct}$, which mediate FCNC involving kaon at one loop.
A more detailed discussion of experimental constraints on 
these couplings is postponed to Sec.~\ref{sec: results}.

\section{Kaon Observables in g2HDM}\label{sec: kaon}
\subsection{ 
Neutral kaon mixing}

{The original measure of CPV in $K \to \pi\pi$, the $\eps$ 
parameter, is rooted in} the complex phase of neutral kaon 
mixing. Defining\footnote{
A different normalization for external states so that
$2m_K M_{12}^\ast = \langle \bar K^0| {\cal H}_{\rm eff}(\Delta  S=2)|K^0\rangle$ is also widely used in the 
literature (see, for example, Ref.~\cite{Buras:1998raa}).
} 
$M_{12}^\ast = \langle \bar K^0| {\cal H}_{\rm eff}(\Delta 
S=2)|K^0\rangle$, one has
\begin{align}
 \eps = \frac{\tilde\kappa_\epsilon e^{i \varphi_\epsilon}}{\sqrt{2}(\Delta M_K)_{\rm exp}}
  ({\rm Im}\, M_{12}^{\rm SM} + {\rm Im}\,M_{12}^{\rm NP}) 
    = e^{i \varphi_\epsilon} (\eps^{\rm SM} + \eps^{\rm NP}),
\label{eq: epsK}
\end{align}
where the phase $\varphi_\epsilon = (43.51 \pm 
0.05)^{\circ}$~\cite{ParticleDataGroup:2020ssz}, and 
correction factor {$\tilde \kappa_\epsilon= 0.94 \pm
0.02$}~\cite{Buras:2008nn} accounts for long distance 
effects. 
Note that {$\eps^{\rm SM, NP}$ are real quantities.}

Experimentally, $\eps$ has been determined with
great accuracy \cite{ParticleDataGroup:2020ssz},
\begin{align}
  |\varepsilon_K| = (2.228 \pm 0.011)\times 10^{-3}.
  \label{eq: epsK-exp}
\end{align}
On the other hand, its prediction in SM is very sensitive 
to $|V_{cb}|$ (the leading SD effect is proportional to 
its 4th power) and therefore the precise determination 
depends on $V_{cb}$ extracted from inclusive or exclusive 
$b\to c \ell \nu$ decays. Detailed analyses of 
$\eps^{\rm SM}$ can be found in Refs.~\cite{UTfit:2006vpt,
Charles:2015gya, Brod:2019rzc}. The recent update in
Ref.~\cite{Brod:2019rzc} finds $\eps^{\rm SM} = 
(2.16 \pm 0.18) \times 10^{-3}$. We impose the 
following~\cite{Aebischer:2020mkv} constraint on NP 
contributions to $\eps$,
\begin{align}
  \eps^{\rm NP} \equiv \kappa \times 10^{-3},
  \quad {\rm where} ~ -0.2 \le \kappa \le 0.2,
  \label{eq: epsK-bound}
\end{align}

On the other hand, $\Delta M_K$ is determined from the 
matrix element $M_{12}$ as,
\begin{align}
  \Delta M_K = 2{\rm Re}\, M_{12}
  \equiv 2[{\rm Re}\, M_{12}^{\rm SM}
       + {\rm Re}\, M_{12}^{\rm NP}].
\end{align}
{While $(\Delta M_K)_{\rm exp} = (5.293 \pm 0.009) 
\times 10^{-3}$~\cite{ParticleDataGroup:2020ssz} is 
precisely measured,} the SM effect is dominated by the 
real part of the box diagrams involving charm quark and 
$W$ exchange. The determination of $\Delta M_K^{\rm SM}$ 
suffers significant uncertainties from QCD corrections 
to the SD part, and from the poorly known LD 
part~\cite{Buras:2014maa}, so we assign a $40\%$ 
uncertainty.

\begin{figure}[b]
\center
\includegraphics[width=.35\textwidth]{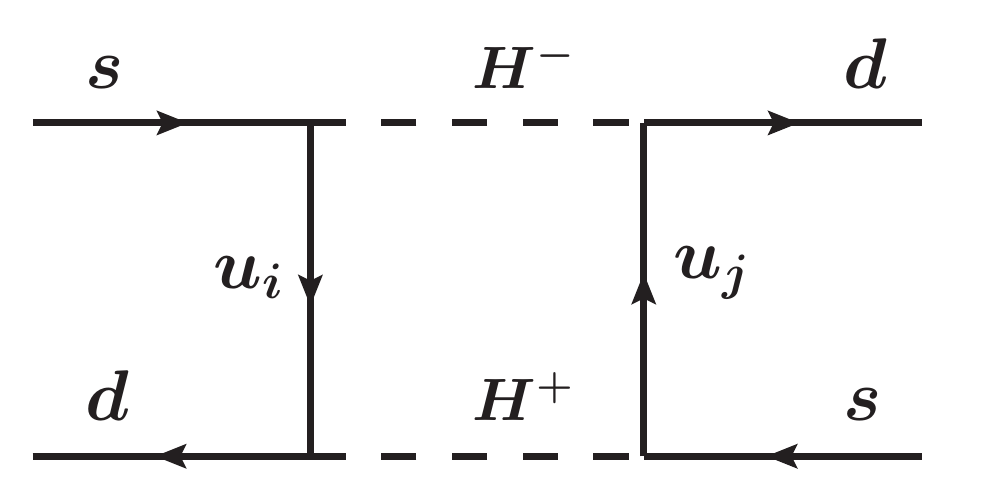}~~~~~
\includegraphics[width=.35\textwidth]{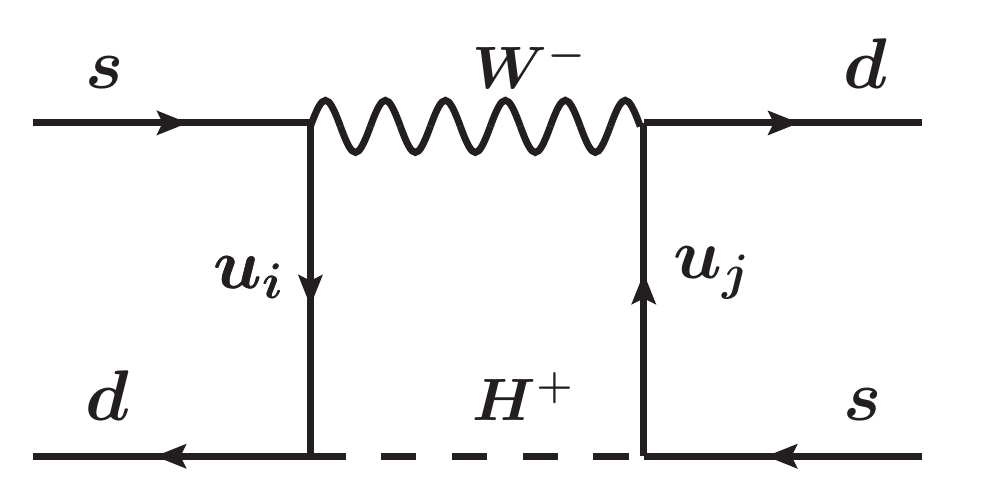}
\caption{
Sample $|\Delta S|=2$ transitions induced by $H^-$ in g2HDM.}
\label{fig: epsK}
\end{figure}

In g2HDM, the NP contribution to {kaon mixing} arise 
from diagrams in Fig.~\ref{fig: epsK}, which generate the 
effective interaction~\cite{Crivellin:2013wna},
\begin{align}
 {\cal H}_{\rm eff} = [C_{HH} + C_{WH}]
  (\bar d \gamma^\mu L s)(\bar d \gamma^\mu L s) + {\rm h.c.}\,,
\label{eq: epsK-Heff}
\end{align}
where $C_{HH}$ is generated by $H^+$--$H^-$ box diagrams,
\begin{align}
  C_{HH} = -\frac{(V_{k1}^\ast \rho_{kj}\rho_{l j}^\ast V_{l 2})
  (V_{m1}^\ast \rho_{mi}\rho_{ni}^\ast V_{n 2})}
  {128 \pi^2 m^2_{H^+}}\,F_{1}(m^2_j/m^2_{H^+}, m^2_i/m^2_{H^+}),
\label{eq: CHH-K}
\end{align}
and $C_{WH}$ is generated by $W^+$--$H^-$ box diagrams,
\begin{align}
  C_{WH} = \frac{g^2\, m_j m_k
  (V_{j1}^\ast \rho_{ij}^\ast V_{i2})
  (V_{l1}^\ast \rho_{lk} V_{k2})}
  {128 \pi^2 m_W^2 m_{H^+}^2} \,
  F_2(m^2_W/m^2_{H^+}, m^2_k/m^2_{H^+}, m^2_j/m^2_{H^+}),
\label{eq: CWH-K}
\end{align}
where $g$ is the weak coupling, and $F_1(x, y)$, 
$F_2(x, y, z)$ are given in Appendix~\ref{app: loop}.
%

\subsection{ 
Kaon direct CPV}

The $\epsp$ parameter
is one of the important kaon observables which probes 
direct CPV in $K \to \pi\pi$ decays. It is convenient to
discuss $K \to \pi\pi$ decays in terms of the isospin
amplitudes $A_{i} = \langle (2\pi)_i| {\cal H}_{eff}(\mu)
| K\rangle$ with $i =1, 2$, and $\mu\sim 1 $ GeV describes 
the physical scale. Then the formula for $\epsp$ can be 
written as~\cite{Cirigliano:2003gt,Buras:2015yba},\footnote{
For brevity of notation, we will use $\epsp$ to denote
$\operatorname{Re}\,(\epsp)$.}
\begin{align}
  {\rm Re}\left(\frac{\varepsilon^\prime}
                     {\varepsilon}\right)
  = -\frac{\omega_+ }{\sqrt{2}|\eps|}
  \left[\frac{\operatorname{Im} A_0}{\operatorname{Re}A_0}(1-\Omega_{\rm eff})
  - \frac{\operatorname{Im} A_2}{\operatorname{Re}A_2}\right],
\end{align}
where $\omega_+\equiv a\, ({\rm Re}\,A_0/{\rm Re}\, A_2)
= (4.53 \pm 0.02) \times 10^{-2}\simeq 1/22$, $a=1.017$, 
and $\Omega_{\rm eff} = (17.0 \pm 9.1) \times 10^{-2}$~\cite{Cirigliano:2019cpi} accounts for 
isospin breaking corrections.
With lattice calculations of amplitudes
$A_2$~\cite{Blum:2015ywa} and $A_0$~\cite{RBC:2020kdj} 
from RBC-UKQCD, one finds in SM,
\begin{align}
  (\epsp)_{\rm SM} \times 10^{4}
      = (21.7 \pm 2.6 \pm 6.2 \pm 5.0) \
 {= (21.7 \pm 8.4) \quad \text{(RBC-UKQCD~2020)}},
\label{eq: epsp-RBC-UKQCD}
\end{align}
where we add uncertainties in quadrature. On the other 
hand, chiral perturbation theory calculation 
gives~\cite{Cirigliano:2019cpi}, 
\begin{align}
  (\epsp)_{\rm SM} = (14 \pm 5) \times 10^{-4}, \quad \text{($\chi$PT 2019)}.
\label{eq: epsp-CHPT}
\end{align}
The current world average for $\epsp$ from NA48 and KTeV 
gives~\cite{NA48:2002tmj, KTeV:2002qqy, KTeV:2010sng},
\begin{align}
(\epsp)_{\rm exp} = (16.6 \pm 2.3) \times 10^{-4},
\end{align}
which is consistent with the theoretical predictions in
Eqs.~\eqref{eq: epsp-RBC-UKQCD} and \eqref{eq: epsp-CHPT}.
\begin{figure}[t]
\center
\includegraphics[width=4.9cm, height=4cm]{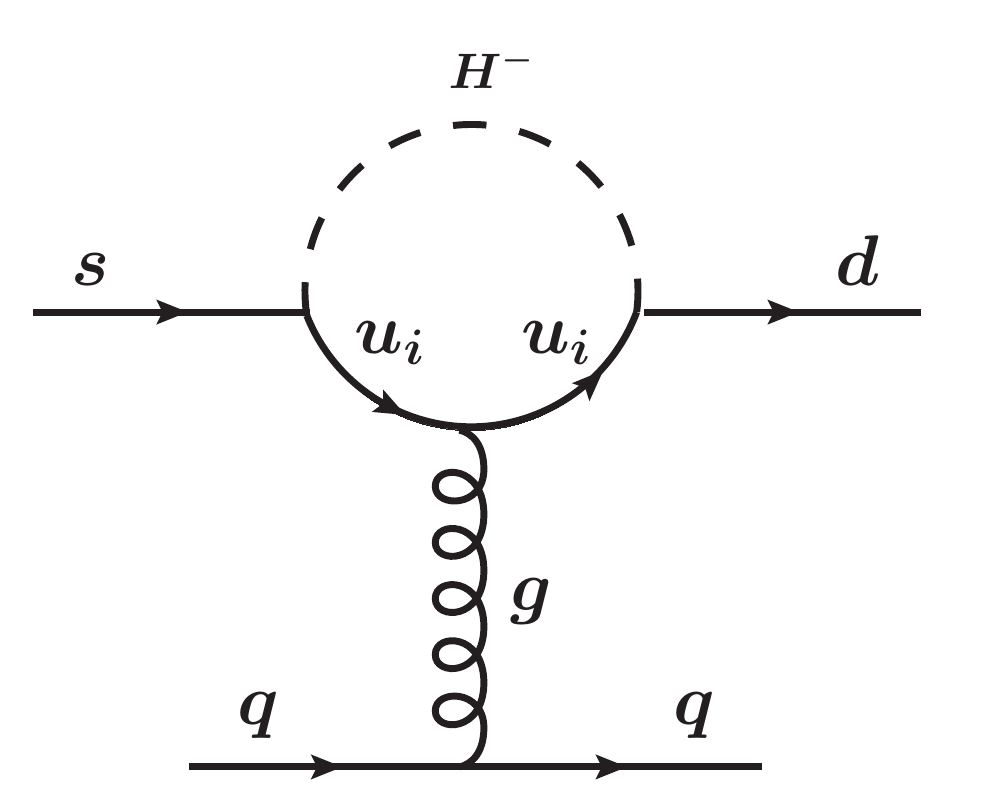}
\includegraphics[width=4.9cm, height=4cm]{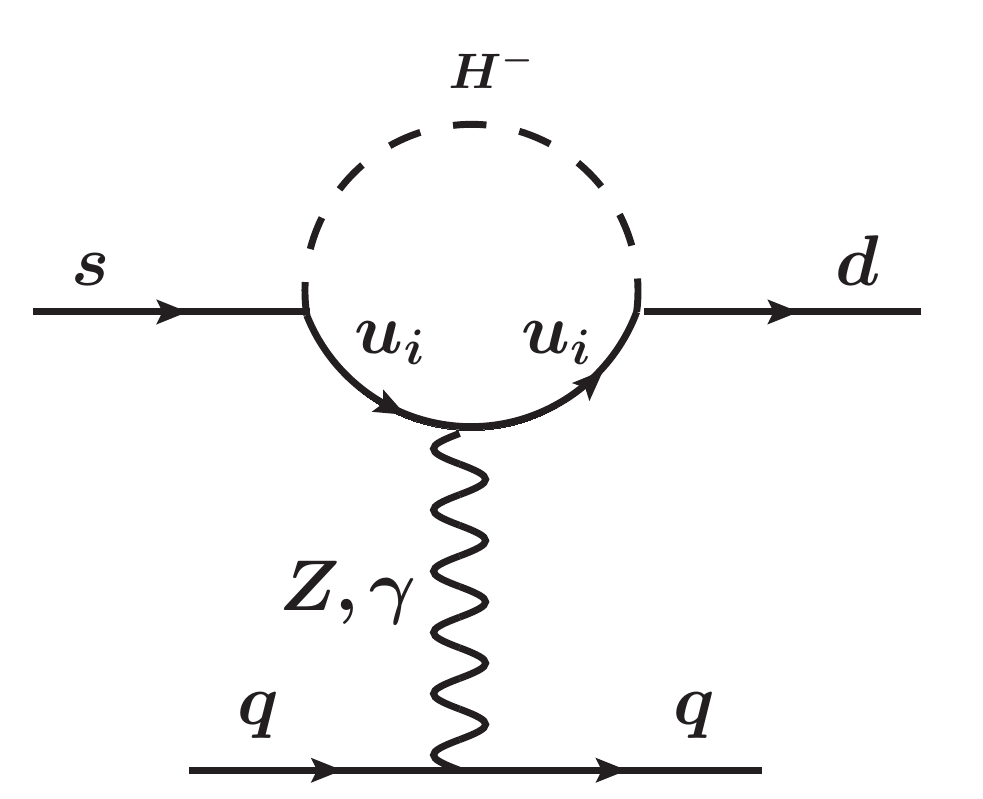}
\includegraphics[width=4.9cm, height=4cm]{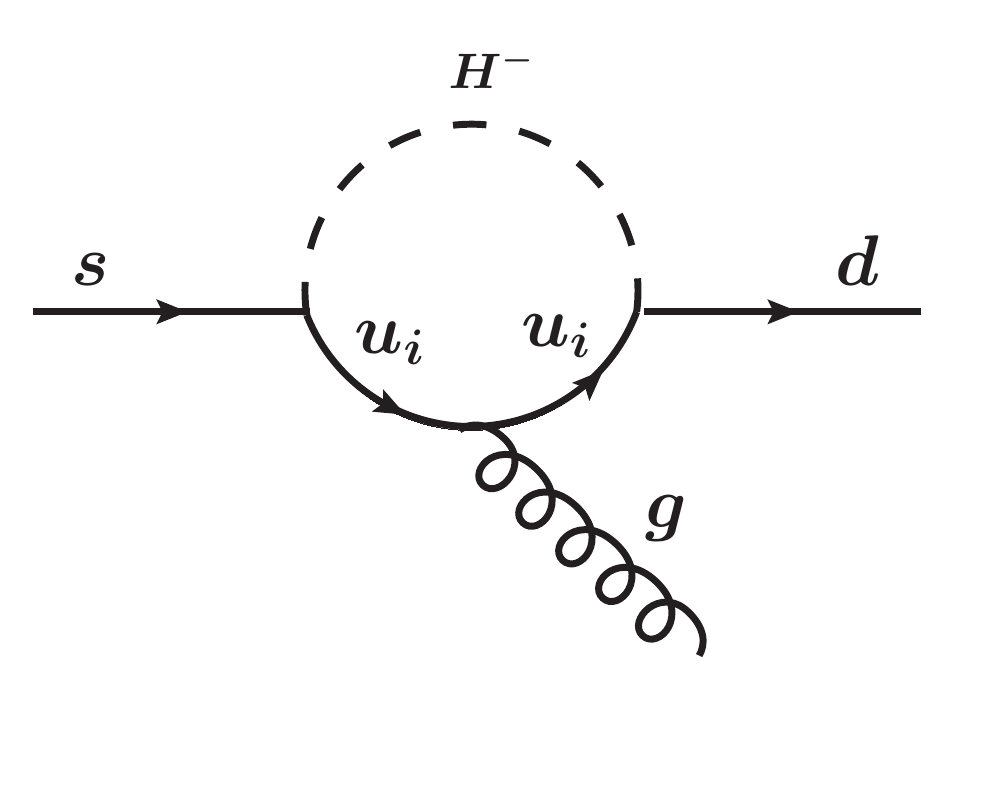}
\caption{
$H^\pm $ induced contributions to $\epsp$ in g2HDM.
Another variant of the second diagram where $Z, \gamma$ propagators couple to $H^-$
is not shown. }
\label{fig: K2pipi}
\end{figure}

The main contribution to $\epsp$ in g2HDM comes from the first 
two diagrams in Fig.~\ref{fig: K2pipi}, which generate the 
following effective interactions,
\begin{align}
  -{\cal H}_{\rm eff} = 
   \sum_{\substack{A = L, R\\{q={u, d}}}}C^q_{VLA}(\bar 
   d\g^\mu L s)(\bar q \g_\mu A q ) \;\ 
&+ \sum_{A = L, R} \tilde C^u_{VLA}(\bar d_\alpha\g^\mu L 
           s_\beta)(\bar u_\beta \g_\mu A u_\alpha )\nn \\ 
&+{~ C_{SLR}^d (\bar d L s) (\bar d R d)} + \, 
       {\rm h.c.}, 
\label{eq: CV-Kpipi}
\end{align}
where we have suppressed color indices for color-singlet operators.

The expressions of coefficients $C^q_{VLA}$ related to 
$s\to d u\bar u $ are as follows: 
\begin{align}
  C^{u}_{VLL} &= -\left[
  g_s^2 G_{\g1}(z_i)
  - 4 e^2 G_{\g 12}(z_i)
  + g^2(-3 + 4 s_W^2)\frac{m_{H^+}^2}{m_W^2}G_Z(z_i)
  \right] \frac{V_{i1}^\ast \rho_{ij}\rho_{k j}^\ast V_{k 2}}
  {96 \pi^2 m^{2}_{H^+}},\\
  C^{u}_{VLR} &= -\left[
  g_s^2 G_{\g1}(z_i)
  - 4 e^2 G_{\g 12}(z_i)
  + 4 g^2s_W^2\frac{m_{H^+}^2}{m_W^2}G_Z(z_i)
  \right] \frac{V_{i1}^\ast \rho_{ij}\rho_{k j}^\ast V_{k 2}}
  {96 \pi^2 m^{2}_{H^+}},\\
  \tilde C^{u}_{VLL} &=  \tilde C^{u}_{VLR}
  =  g_s^2 \frac{V_{i1}^\ast \rho_{ij}\rho_{k j}^\ast V_{k 2}}
  {32 \pi^2 m^{2}_{H^+}} G_{\g 1}(z_i),
\end{align}
while those related to $s\to d d\bar d$ are given by,
\begin{align}
  C^{d}_{VLL} &= - \left[
  - 2 g_s^2 G_{\g1}(z_i)
  + 2 e^2 G_{\g 12}(z_i)
  + g^2(3 -2 s_W^2)\frac{m_{H^+}^2}{m_W^2}G_Z(z_i)
  \right] \frac{V_{i1}^\ast \rho_{ij}\rho_{k j}^\ast V_{k 2}}
  {96 \pi^2 m^{2}_{H^+}},\\
  C^{d}_{VLR} &= -\left[
   g_s^2 G_{\g1}(z_i)
  + 2 e^2 G_{\g 12}(z_i)
  - 2 g^2 s_W^2\frac{m_{H^+}^2}{m_W^2}G_Z(z_i)
  \right] \frac{V_{i1}^\ast \rho_{ij}\rho_{k j}^\ast V_{k 2}}
  {96 \pi^2 m^{2}_{H^+}},\\
  {C^{d}_{SLR}} &= -2 \tilde C^{u}_{VLR}.
\end{align}
Here $g_s$, $e$, $g$ arise from $g$-, $\gamma$- and 
$Z$-penguin diagrams, respectively, and repeated indices
are summed over. The loop functions $G_{\g 1, \g12, Z}(x)$
are listed in Appendix~\ref{app: loop}.

The last diagram in Fig.~\ref{fig: K2pipi} give rise to the
chromo-magnetic dipole interaction:
\begin{align}
  {\cal H}_{\rm eff}
  = -C_{8g} m_s(\bar d \sigma_{\mu\nu}T^a R\, s) G_{\mu\nu}^a
    + {\rm h.c.},
  \label{eq: C8g-Kpipi}
\end{align}
where coefficient $C_{8g}$ is
\begin{align}
  C_{8g} =  g_s \frac{V_{i1}^\ast \rho_{ij}\rho_{k j}^\ast V_{k 2}}{32 \pi^2 m^{2}_{H^+}} F_{\sigma_1}(m^2_i/m^2_{H^+})\,,
\end{align}
where the loop function $F_{\sigma_1}(x)$ is given in 
Appendix~\ref{app: loop}. But we note that the dipole 
operator contribution in our case is very small compared 
to those from Eq.~\eqref{eq: CV-Kpipi} and can be ignored. 

The NP contribution to $(\epsp)_{\rm NP}$ can be 
calculated from the following formula~\cite{Aebischer:2020jto}:\footnote{
Note that ${\cal H}_{\rm eff}$ in 
Ref.~\cite{Aebischer:2020jto} has the form $\sum_i 
C_i/(1\,{\rm TeV})^2 (\bar s \Gamma_i d) 
(\bar q \Gamma_i q)$, while we have $\sum_i C_i (\bar d 
\Gamma_i s) (\bar q \Gamma_i q)$. Therefore, 
Eq.~\eqref{eq: epsp-NP} contains complex-conjugate Wilson 
coefficients rescaled by $(1\, {\rm TeV})^2$ compared to 
the similar expression given in 
Ref.~\cite{Aebischer:2020jto}.
}
\begin{align}
 {(\epsp)_{\rm NP}
   = \sum_i P_i(\mu_{\rm EW}) {\rm Im}[C_i^\ast (\mu_{\rm EW})
                              - C_i^{\prime \ast} (\mu_{\rm EW})]
                                \times (1\ {\rm TeV})^2,}
\label{eq: epsp-NP}
\end{align}
where $\mu_{\rm EW} \simeq 160$~GeV corresponds to the 
electroweak scale, and the $P_i$ factors capture 
information of hadronic matrix elements 
$\langle(\pi\pi)_{I}|O_iK\rangle$. The numerical values 
of $P_i$, evaluated using matrix elements provided by RBC-UKQCD~\cite{Blum:2015ywa,RBC:2020kdj}, are given in 
Ref.~\cite{Aebischer:2020jto}. Since the Wilson coefficients $C_i^{(\prime)}$ in Eq.~(\ref{eq: epsp-NP}) are defined at 
the electroweak scale, but those in Eqs.~\eqref{eq: CV-Kpipi}
and \eqref{eq: C8g-Kpipi} are at NP scale $\mu_{\rm NP}$, 
one needs to evolve $C_i^{(\prime)}(\mu_{\rm NP})$ down to 
the electroweak scale using renormalization group equations 
(RGE) before using the NP formula given in Eq.~\eqref{eq: 
epsp-NP}. Note also that though operators with $q=s, c, b$ 
in Eq.~\eqref{eq: CV-Kpipi} do not arise at the $K \to 
\pi\pi$ factorization scale, they can contribute through 
RGE above this scale~\cite{Aebischer:2018rrz}. But as their 
contribution turns out to be very small compared to the 
ones with $q=u,d$~\cite{Aebischer:2020jto}, we ignore these 
operators for simplicity.

As already stated, the latest SM prediction for $\epsp$ agrees well 
with the experimental value. But the uncertainties in both values, 
especially theory, are quite large at present, which allows the
following range of values from NP contributions~\cite{Aebischer:2020jto},
\begin{align}
  -4 \times 10^{-4} \lesssim (\epsp)_{\rm NP} \lesssim 10 \times 10^{-4}.
  \label{eq: epsp-bound}
\end{align}

\subsection{ 
 $\Kpnn$ amd $\KLnn$}
The branching ratio of $\Kpnn$ in the SM is given as \cite{Buras:2015qea},
\begin{align}
  {\cal B}(\Kpnn) = \kappa_{+}(1+\delta_{\rm EM})
\left\{\left(\frac{\operatorname{Im}{\left[v_t\,X(x_t)\right]}}
                    {\l^5} \right)^2
  + \left(\frac{\operatorname{Re} {[v_c]}}{\l} P_c
              + \frac{\operatorname{Re}{\left[v_t\,X(x_t)\right]}}
                     {\l^5} 
    \right)^2\right\}.
\label{eq: Kpnn}
\end{align}
Here $\l \equiv |V_{us}|$, $v_i \equiv V_{is}^\ast V_{id}$
are CKM factors, $\delta_{\rm EM}=-0.003$ accounts for 
radiative corrections to the decay, and the factor 
$\kappa_{+}  = (5.173 \pm 0.025)(\l/0.025)^8 \times 
10^{-11}$~\cite{Buras:2015qea} contains information about 
FCNC hadronic matrix elements obtained from semileptonic 
decays of kaons~\cite{Mescia:2007kn}. The loop function 
$X(x_t) = 1.462 \pm 0.017$ \cite{Brod:2021hsj}, where 
$x_t=m_t^2/m_W^2$, contains pure SD from top quark while 
$P_c=(0.405 \pm 0.024)(0.225/\l)^4$~\cite{Buras:2005gr,  Buras:2006gb, Brod:2008ss, Isidori:2005xm, Buras:2021nns}
includes both SD and LD contributions from charm quark. 

{For the SM branching ratio of $\Kpnn$, we obtain,
\begin{align}
  {\cal B}(\Kpnn)_{\rm SM} = (9.07 \pm 0.82)\times 10^{-11},
  \label{eq: Kpnn-SM-value}
\end{align}
which is consistent with the commonly cited value of Ref.~\cite{Buras:2015qea},}
whereas a more recent analysis \cite{Brod:2021hsj} finds a lower 
but more precise value of $(7.7 \pm 0.6)\times 10^{-11}$ in SM. 
The uncertainties in SM are dominated by the CKM parameters 
$V_{cb}$ and phase angle $\gamma$ \cite{Buras:2015qea}.

On the experimental side, NA62 has reported \cite{NA62:2021zjw},
\begin{align}
  {\cal B}(\Kpnn)_{\rm exp} = (10.6^{+4.0}_{-3.4} \pm 0.9)\times 10^{-11},
  \label{eq: Kpnn-NA62}
\end{align}
where the first error is statistical and the second systematic.
The measurement is based on data collected during the 2016-2018 
runs and {improves} the previous measurement by E949 at
Brookhaven~\cite{BNL-E949:2009dza}. Eq.~\eqref{eq: Kpnn-NA62}
agrees within $1\sigma$ of Eq.~\eqref{eq: Kpnn-SM-value}, but
the large statistical error still allows significant NP
contribution. NA62 will continue collecting data in the next 
few years, and aims at measuring the branching ratio to $10\%$
precision by 2024~\cite{NA62:2020upd}. 

\begin{figure}[t]
\center
\includegraphics[width=4.9cm, height=4cm]{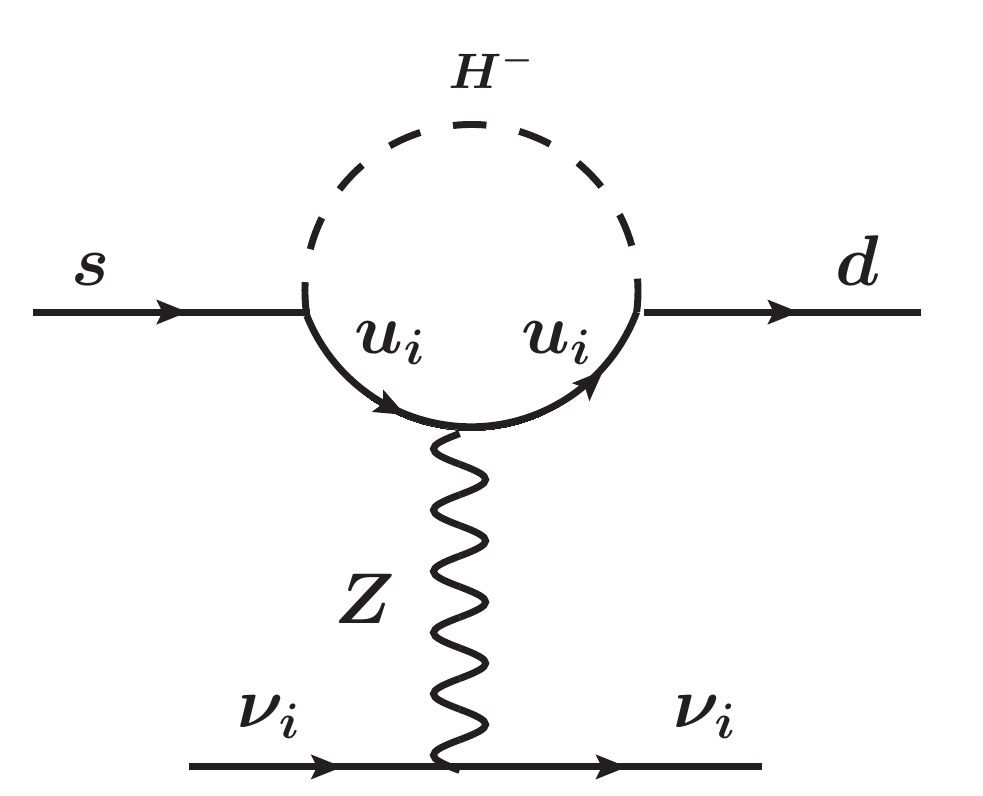}
\includegraphics[width=4.9cm, height=4cm]{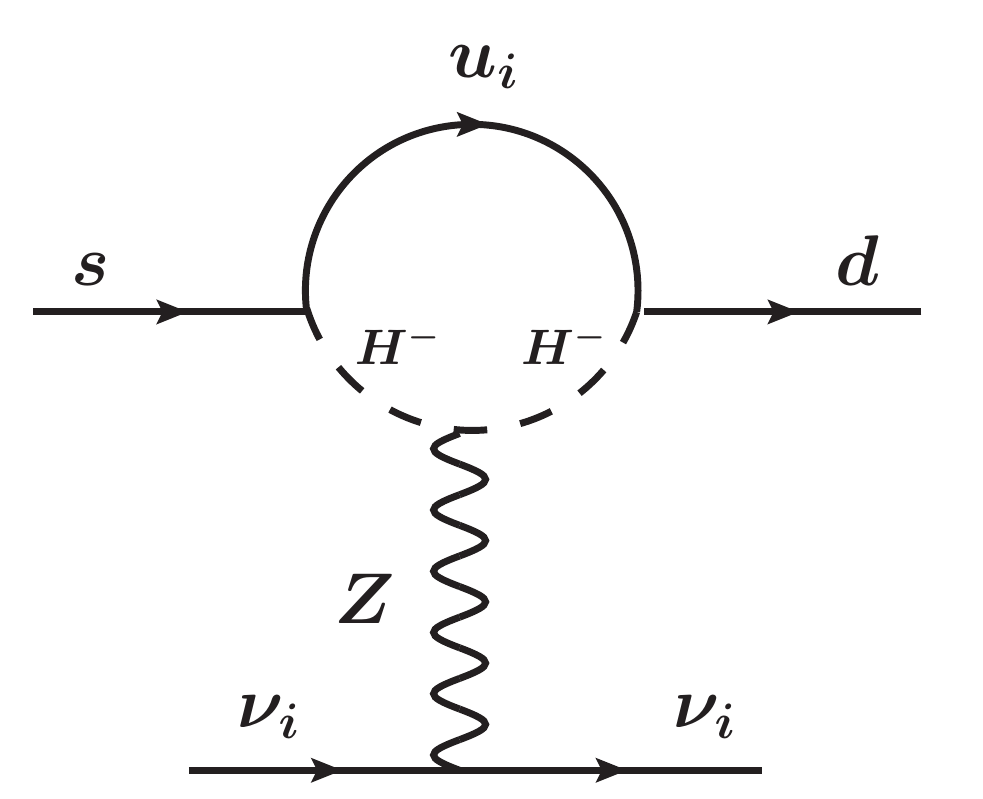}
\caption{
$H^-$ induced $Z$-penguin contributions to $s\to d \nu \bar \nu$
in g2HDM.
}
\label{fig: s2dnn}
\end{figure}

In g2HDM, the $Z$-penguin diagrams with $H^\pm$ in the loop 
as shown in Fig.~\ref{fig: s2dnn}, generate a purely  
left-handed effective interaction as in SM,
\begin{align}
{\cal H}_{\rm eff}= \frac{4\,G_F}{\sqrt{2}}C_{LL}^{a, b}\,
    \left(\bar s \g_\mu L d\right)
    \left(\bar \nu_a \g^\mu L \nu_b\right) + {\rm h.c.},
\label{eq: b2snunu-Zpenguin}
\end{align}
where $a, b$ denote neutrino flavors. The Wilson coefficient 
$C_{LL}^{a, b}$ is\footnote{
Our result for $C_{LL}^{a, b}$ in Eq.~\eqref{eq: b2snunu-WC}
{differs from Ref.~\cite{Iguro:2019zlc} by {a factor of 2;} checking with the authors, they agree with our formula.
Our result agrees with the corresponding expression of
Ref.~\cite{Geng:1988bq}, but these authors took the limit of
diagonal $\rho^u$, which differs from our case}.}
\begin{align}
C_{LL}^{a, b} = -\frac{\delta_{ab}}{16\pi^2}
   (V^\dagger \r^u)_{2i}(\r^{u\dagger} V)_{i1} 
                 \,G_Z(m_i^2/m^2_{H^+}).
\label{eq: b2snunu-WC}
\end{align}
where the loop function $G_Z$ is given in 
Appendix~\ref{app: loop}. 

The g2HDM modification to SM branching ratio is effected by the 
simple replacement,
{
\begin{align}
  X(x_t) \to X_{\rm eff} \equiv  X(x_t)
  + \frac{2 \pi s_W^2 }{ \alpha v_t} C_{LL}^{a, b},
\label{eq: XNP}
\end{align}
}
where $s_W$ is the Weinberg angle and $\alpha$ the fine
structure constant.

%


For the CP-violating decay $\KLnn$, the SM result
is~\cite{Buras:2015qea},
\begin{align}
  {\cal B}(\KLnn) = \kappa_{L}
\left(\frac{\operatorname{Im} \left[v_t X(x_t)\right]}{\l^5}
     \right)^2,
\label{eq: KLnn}
\end{align}
where $\kappa_{L} = (2.231\pm 0.013)(\l/0.025)^8\times 10^{-10}$,
with remaining parameters defined already. Eq.~\eqref{eq: KLnn} 
is completely dominated by SD top contribution with charm  contribution negligible,
leading to a more accurate prediction of branching ratio. {Within SM, we obtain
\begin{align}
  {\cal B}(\KLnn)_{\rm SM} = (3.24 \pm 0.36) \times 10^{-11}\,,
\end{align}
consistent with commonly cited result of Ref.~\cite{Buras:2015qea},}
where theory uncertainties are again dominated by CKM parameters,
with $|V_{ub}|$ contributing the most \cite{Buras:2015qea}. 
{The analysis in Ref.~\cite{Brod:2021hsj} obtains a slightly
lower central value ${\cal B}(\KLnn)_{\rm SM} = (2.59 \pm 0.29)
\times 10^{-11}$.}

{The KOTO experiment at J-PARC sets a $90\%$ C.L.
bound~\cite{KOTO:2020prk} of
\begin{align}
  {\cal B}(\KLnn)_{\rm exp} < 4.9 \times 10^{-9},
  \label{eq: KOTO2021}
\end{align}
which is based on data collected during 2016-2018, where three
candidate events were observed in the signal region. KOTO
subsequently uncovered contamination from $K^\pm$ and 
scattered $K_L$ decays, giving a total number of $1.22 \pm 0.26$
background events~\cite{KOTO:2020prk}. Having identified the
potential source of the three observed events, it led to a 
weaker bound than the single event sensitivity. A previous
analysis based on a {\it smaller} data set collected in 2015 
gave a slightly better bound of ${\cal B}(\KLnn)_{\rm exp} 
< 3.0\times 10^{-9}$~\cite{KOTO:2018dsc}, in good part because
no signal events were observed.
}

The future prospect for the measurement of $\KLnn$ is 
{rather good}. KOTO will resume data collection 
and expects to measure the decay at SM sensitivity in 
the next few years~\cite{Shiomi:2021oht}. A proposed 
experiment KLEVER~\cite{KLEVERProject:2019aks} at CERN 
plans to measure the decay at $\sim20\%$ precision. There 
is also the proposal for KOTO Step-2~\cite{Nomura:2020oor, 
Aoki:2021cqa}, aiming at single event sensitive of 
$\o (10^{-13})$ to measure the decay rate.

The g2HDM modification of the SM branching ratio, 
Eq.~(\ref{eq: KLnn}), is achieved by replacing $X(x_t)$ 
according to Eq.~\eqref{eq: XNP}, just as for $\Kpnn$.

\subsection{ 
$K_{L,S} \to \mu^+\mu^-$}

With effective hamiltonian for $K^0 \to \mu^+\mu^-$ defined as,
\begin{align}
  {\cal H}_{\rm eff} =
  - C_A (\bar s \g^\mu P_L d)(\mu \g_\mu\g_5 \mu)
  + {\rm h.c.}\,,
\label{eq: Heff-KLmm}
\end{align}
the branching ratio for $\KLmm$ is given 
by~\cite{Chobanova:2017rkj},
\begin{align}
  {\cal B}(\KLmm) = \tau_L\frac{f_K^2 m_K m_\mu^2\beta _\mu }{4 \pi }
  \left| \operatorname{Re}(C_A) -  \frac{G_F^2 m_W^2}{\pi^2}A^\mu_{L\g\g}\right|^2,
\label{eq: brKLmm}
\end{align}
where the first term corresponds to the SD contribution 
and the second represents LD contributions, with $\tau_{L}$ 
the $K_{L}$ lifetime~\cite{ParticleDataGroup:2020ssz}, 
$f_K$ the decay constant~\cite{Aoki:2021kgd}, and 
$\beta_\mu=$ {\small$\sqrt{1-4m_\mu^2/m_K^2}$}. Within SM, 
$C_A^{\rm SM} = -(G_F^2 m_W^2/\pi^2) (v_t Y_t + v_c Y_c)$, 
where $Y_t = 0.950\pm 0.049$ and $Y_c = (2.95 \pm 0.46) 
\times 10^{-4}$ denote contributions from top and charm 
quarks, respectively~\cite{Gorbahn:2006bm}, while the LD 
contributions have been studied in Refs.~\cite{Ecker:1991ru, Isidori:2003ts, DAmbrosio:2017klp, Mescia:2006jd}. 

{The numerical value of $A^\mu_{L\g\g}$ can be found 
in Ref.~\cite{Chobanova:2017rkj}, but the sign of 
$A^\mu_{L\g\g}$ is not known, which can be constructive 
($-$) or destructive ($+$) with the SD contribution.} 
The corresponding predictions in the SM
are~\cite{Ecker:1991ru, Isidori:2003ts, Gorbahn:2006bm, DAmbrosio:2017klp},
\begin{align}
  {\cal B}(\KLmm)_{\rm SM}
  = \left\{\begin{aligned}
  (8.11 \pm 0.49 \pm 0.13) \times 10^{-9}, & \quad ({\rm for}\,-{\rm sign})\\
  (6.85 \pm 0.80 \pm 0.06) \times 10^{-9}, & \quad ({\rm for}\,+{\rm sign})
  \end{aligned} \right.
\label{eq: KLmm-SM}
\end{align}
where the first uncertainty is from LD contribution, 
while the second contains parametric uncertainties from 
e.g. CKM elements. A precise measurement of $K_{L,S} \to 
\mu^+\mu^-$ interference can 
help~\cite{DAmbrosio:2017klp} determine the sign of 
$A^\mu_{L\g\g}$.
   
Based on Refs.~\cite{E871:2000wvm, Akagi:1994bb, 
E791:1994xxb}, the world average for ${\cal B}(\KLmm)$ 
is~\cite{ParticleDataGroup:2020ssz},
\begin{align}
  {\cal B}(\KLmm)_{\rm exp} = (6.84 \pm 0.11) \times 10^{-9}
  \quad ({\rm PDG~2020}),
  \label{eq: KLmm-exp}
\end{align}
{which is measured to 1.6\% precision and seem to favor
destructive interference from LD effect.} The g2HDM 
contributions to $\KLmm$ are dominated by the same 
diagrams as in Fig.~\ref{fig: s2dnn}, but with $Z\nu\nu$ 
vertex replaced by $Z\mu\mu$. The $\g$-penguin diagrams 
are absent due to a Ward identify for on-shell leptons. 
The corresponding contribution to $C_A$ is,
\begin{align}
  C_A^{\rm NP}
  = \frac{G_F}{8\sqrt{2}\pi^2}V_{i2}^\ast
    \rho_{ij}\rho_{k j}^\ast V_{k 1}G_Z(x_i).
\end{align} 
The $K^0 \to e^+e^-$ decay is suppressed by $m_e^2/m_\mu^2$ compared with dimuons.


The $\KSmm$ rate depends on the imaginary part of SD 
contributions, hence a sensitive probe of NP with complex phases.
The branching ratio is~\cite{DAmbrosio:2017klp},
\begin{align}
  {\cal B}(\KSmm)
  =  \tau_S\frac{f_K^2 m_K m_\mu^2\beta _\mu }{4 \pi }
   \left[ \left(\operatorname{Im}(C_A)\right)^2
        + \left|\frac{\beta_\mu G_F^2m_W^2}{\pi^2}
                   B^\mu_{S\g\g}\right|^2\right],
\label{eq: brKSmm}
\end{align}
with $C_A$ as defined in Eq.~\eqref{eq: Heff-KLmm}, and 
$B^\mu_{S\g\g}$ arises~\cite{DAmbrosio:2017klp} from LD 
effects, which only add in quadrature to SD. The rate is 
suppressed by the $K_S$ lifetime, $\tau_S$, down to~\cite{Ecker:1991ru, Isidori:2003ts, Gorbahn:2006bm, 
DAmbrosio:2017klp},
\begin{align}
  {\cal B}(\KSmm)_{\rm SM}
  = (4.99_{\rm LD} + 0.19_{SD})\times 10^{-12}
  = (5.2 \pm 1.5) \times 10^{-12}.
\end{align}
The current upper limit by LHCb~\cite{LHCb:2020ycd} at 
$90\%$ C.L. {is ${\cal B}(\KSmm) < 2.1 \times 10^{-10}$,} which improves their previous 
bound~\cite{LHCb:2017qna} by a factor of four. LHCb 
Upgrade~II will improve the limit to below 
$\o(10^{-11})$, and should {approach} SM 
sensitivity~\cite{Cerri:2018ypt}.

\section{Results}\label{sec: results}

For our numerical study, relevant parameters are the 
complex extra Yukawa couplings $\rtt$, $\rtc$, $\rct$, 
and the charged Higgs mass $m_{H^+}$. We use the software  
package \texttt{Flavio}~\cite{Straub:2018kue}; all Wilson 
coefficients are adapted to \emph{flavio  basis}~\cite{Aebischer:2017ugx}, then QCD-evolved from 
high scale to the relevant process scale using the package 
\texttt{Wilson}~\cite{Aebischer:2018bkb}. For extra Higgs 
boson masses in g2HDM, sub-TeV values are favored for 
strongly first order electroweak phase transitions in 
the early Universe~\cite{Kanemura:2004ch, Fuyuto:2017ewj, Fuyuto:2019svr}. But as we will see later, rare kaon decays 
offer excellent probes of {\it heavier} $H^+$. Therefore, 
we give results for $m_{H^+} =$ 400 and 1000 GeV, and refer 
to the former as light $H^+$ scenario, while the latter is  
called the heavy $H^+$ scenario.

\begin{table}[b]
\centering
\begin{tabular}{|c|r|}
\hline
Observable & 
 Measurement   \\ 
\hline\hline
$\Delta M_{B_s}$ & $(17.741 \pm 0.020)~{\rm ps}^{-1}$~\cite{ParticleDataGroup:2020ssz}\\
$\Delta M_{B_d}$ & $(0.5065 \pm 0.0019)~{\rm ps}^{-1}$~\cite{ParticleDataGroup:2020ssz}\\
$S_{\psi K_S }$ & $(0.699 \pm 0.017)$~\cite{HFLAV:2019otj} \\
$S_{\psi \phi}$ & $(0.050 \pm 0.019)$~\cite{HFLAV:2019otj} \\
${\cal B}(B \to X_s \gamma)$ & $(3.32 \pm 0.15) \times 10^{-4}$~\cite{HFLAV:2019otj} \\
${\cal B}(B_s \to \mu^+\mu^-)$ & $(3.09^{+0.46 \,+ 0.15}_{-0.43\,-0.11})
\times 10^{-9}$~\cite{LHCb:2021awg} \\
\hline
\hline
\end{tabular}
\caption{Experimental data of various $B$ meson observables.}
\label{tab: B-meson-data}
\end{table}
%

\subsection{ 
{Effect of B sector and 
  kaon CPV constraints}}

We first consider constraints from B sector. We list the 
$B$ meson observables considered and the corresponding 
measurements in Table~\ref{tab: B-meson-data}. The g2HDM 
{formulas} for $B \to X_s \gamma$ and $B_s \to 
\mu^+\mu^-$ decay branching ratios can be found in many 
works (see, for example, Refs.~\cite{Iguro:2017ysu, Crivellin:2019dun, Athron:2021auq}).
The discussion of 
neutral $B_q$ $(q=s, d)$ mass difference $\Delta M_{B_{q}}$ 
and mixing-induced CP asymmetries $S_{\psi K_S}$ and 
$S_{\psi \phi}$ as probes of $B_q$-mixing phases from 
$B_d \to \psi K_S$ and $B_s \to \psi \phi$, respectively, 
is relegated to Appendix~\ref{app: acp}.

\begin{figure}[t]
\center
\includegraphics[width=6cm,height=5.5cm]{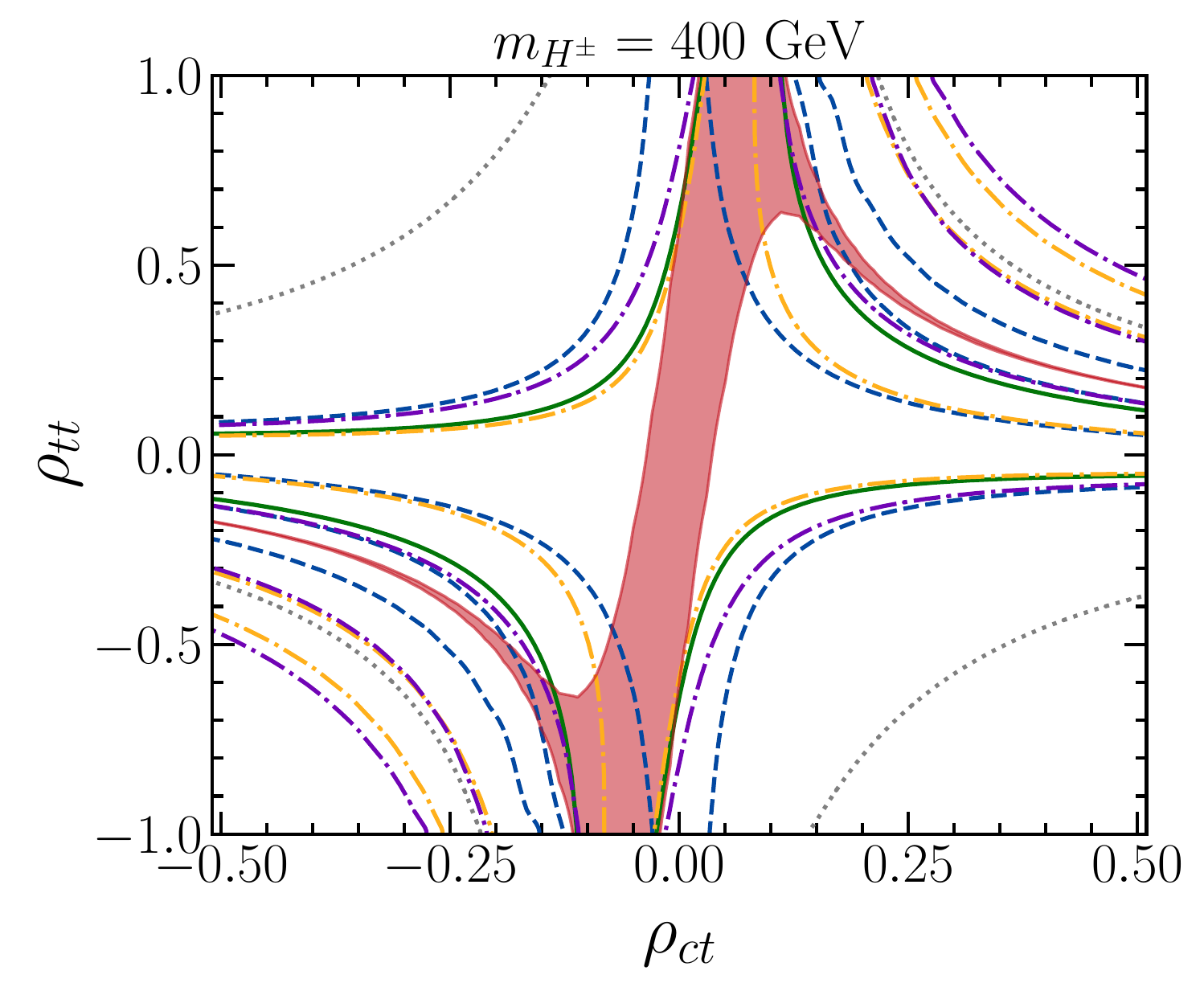}
~~~\includegraphics[width=8cm,height=5.5cm]{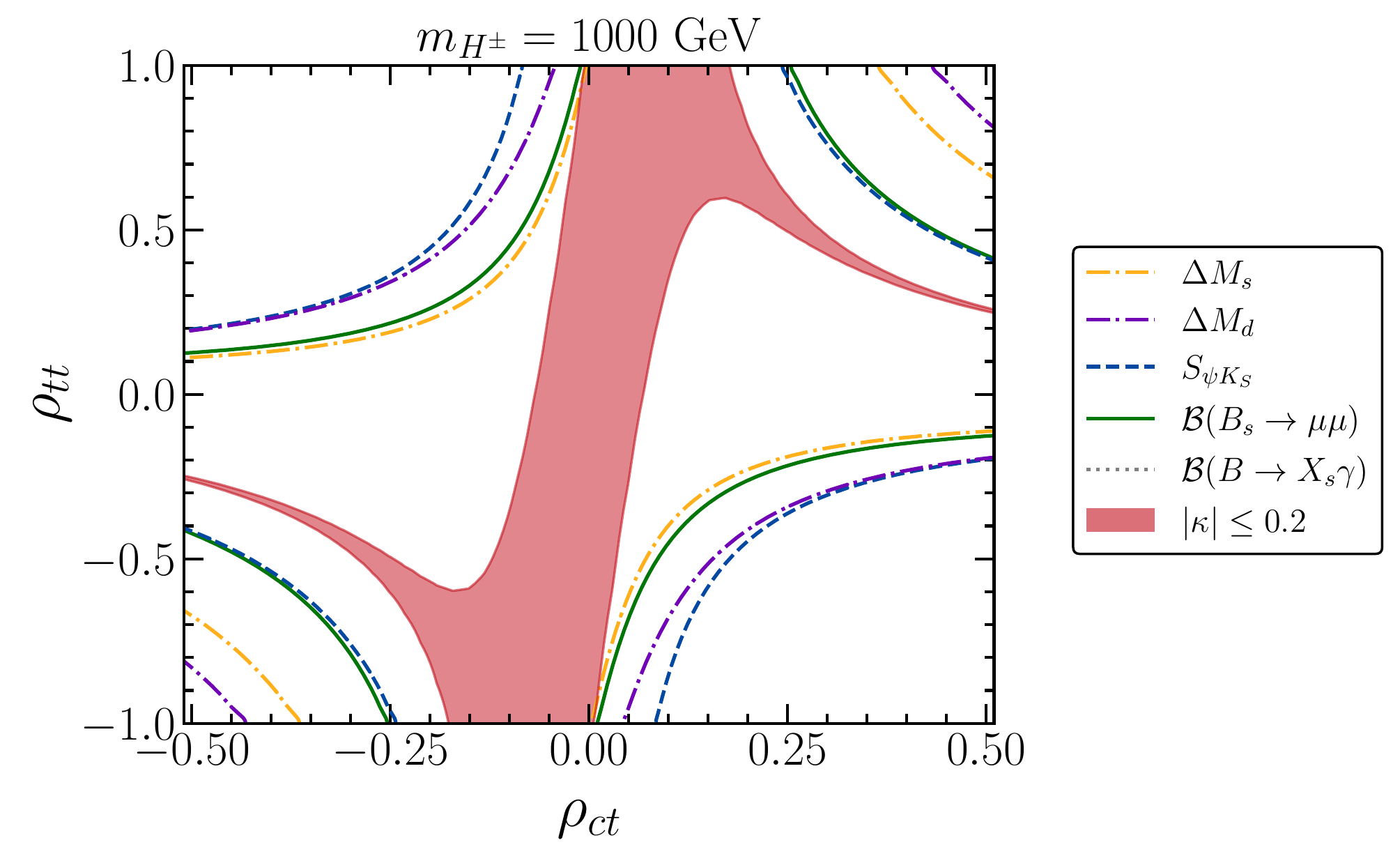}
\caption{
B sector and $\eps$ constraints in $\r_{ct}$--$\r_{tt}$ 
plane for $m_{H^+} = 400$ and 1000 GeV. Colored lines 
indicate $2\sigma$ range of experimental data, outside of 
which are \emph{ruled out}. The filled red band corresponds  to the region \emph{allowed} by $\eps^{\rm NP}$.
}
\label{fig: rct-rtt-B}
\end{figure}

In Fig.~\ref{fig: rct-rtt-B}, we illustrate  $2\sigma$ constraints from 
$B$ observables in the $\rct$--$\rtt$ plane, where the relatively 
weak bound from  $S_{\psi \phi}$ is not shown. 
To contrast with B sector, we show also the region (red) allowed 
by $\eps$ using Eq.~\eqref{eq: epsK-bound} (recall $\eps^{\rm NP} 
= \kappa \times 10^{-3}$). For light $m_{H^+} = 400$ GeV, the B 
sector rules out significant portions of parameter space, but 
still allows large values of $\rtt$ and $\rct$, especially along 
the axes, i.e. in regions where one of the couplings is vanishing. 
But as seen from Fig.~\ref{fig: rct-rtt-B}, $\eps$ provides the 
most severe constraint. Combined with the B sector, it essentially 
rules out a sizable $\rct$, but $\rtt$ can still be quite large 
since $\eps$ has weaker sensitivity to $\rtt$ compared with 
$\rct$. However, for heavy $m_{H^+}=$ 1000 GeV, the B sector 
bounds weaken while the region allowed by $\eps$ broadens. But 
now $\eps$ truly becomes the leading constraint in the whole 
parameter space. From Fig.~\ref{fig: rct-rtt-B}, one also notes 
that contributions from $\rtt$ and $\rct$ to $\eps$ (also to 
$\Delta M_{B_q}$ and $S_{\psi K_S}$) tend to cancel each other, 
as reflected in the narrow red bands for large $\rct$ in heavy 
$H^+$ case.

In Fig.~\ref{fig: rct-rtt-B}, $\rct$ and $\rtt$ are taken as 
real. But the essence of Yukawa couplings are their 
complexity. Allowing complex phases, the cancellation region 
hence allowed parameter space changes. To explore g2HDM 
effects in the kaon sector, we perform a parameter scan that 
takes the phases of $\phi_{ij}\equiv \arg\rho_{ij}$ into 
account. Specifically, we scan over:
\begin{align}
  |\rtt|,\,|\rtc| \in [0, 1],\  
   \phi_{tt},\,\phi_{tc} \in [-\pi, \pi ]; \quad 
  \quad\, |\rct| \in [0, 0.3],\ \,\phi_{ct} \in [-\pi, \pi ].
\label{eq: scan}
\end{align}
Each parameter is varied uniformly to generate a sample size
of {a quarter million random points}. The smaller range of 
$|\rct|$ is chosen from hindsight, that combined flavor 
constraints on $\rct$ will rule out values larger than 
$\sim 0.2$, as will be shown later. Fixing to narrower range 
also means we can obtain a denser population of allowed points.

\begin{figure}[t]
\center
\includegraphics[width=0.49\textwidth]{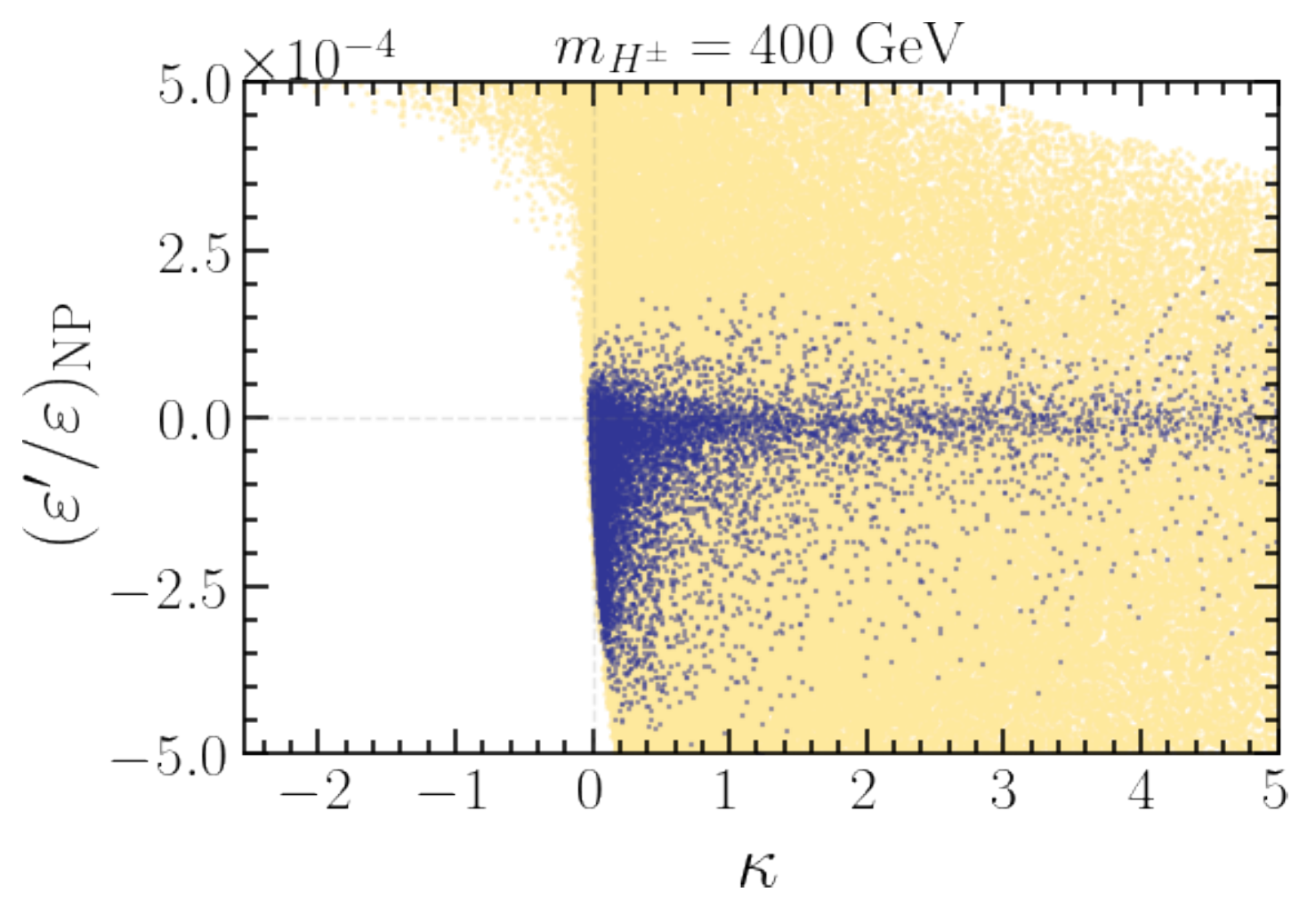}
\includegraphics[width=0.49\textwidth]{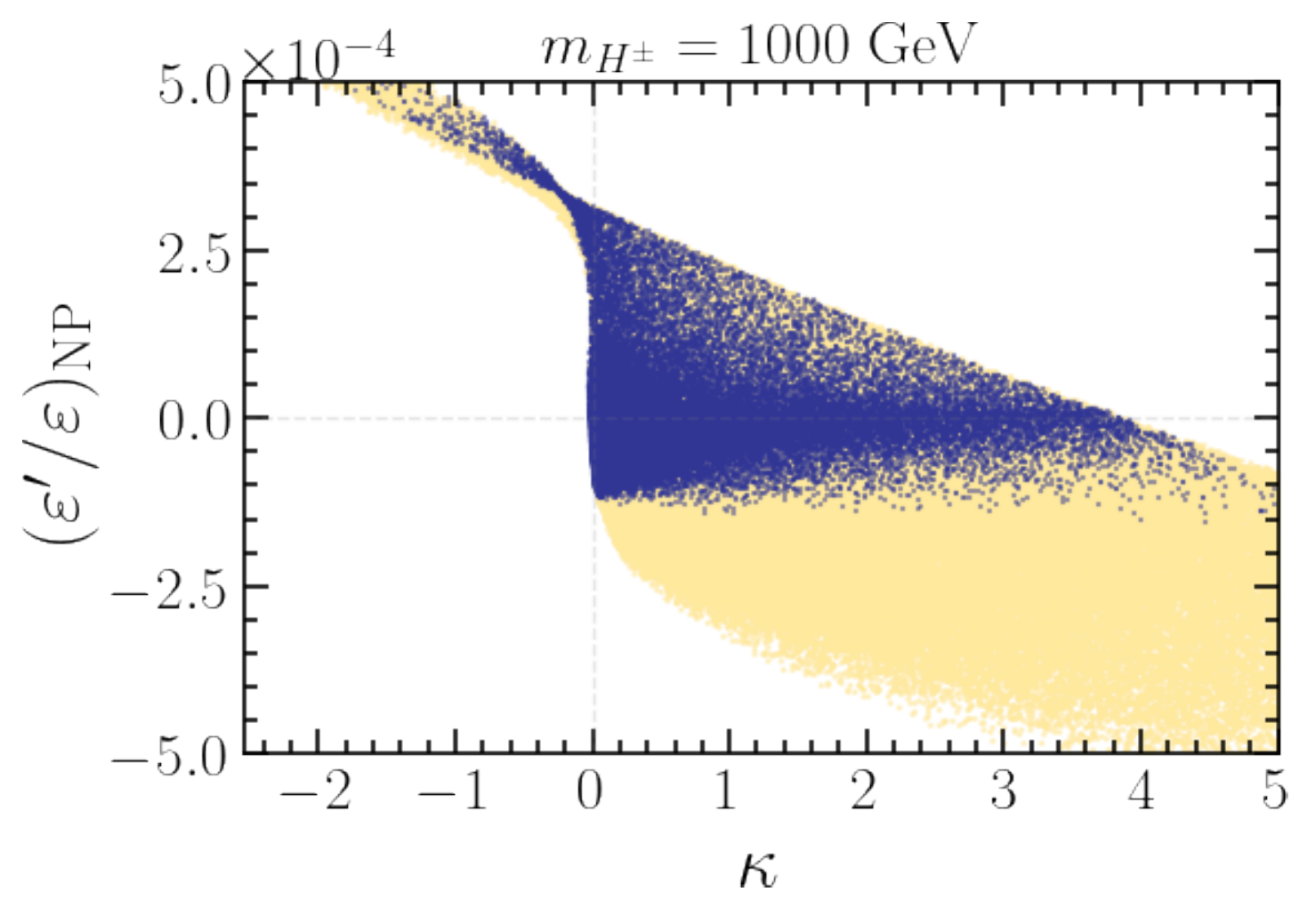}
\caption{ 
Scatter plot showing range of values accessible in g2HDM 
for $\eps^{\rm NP}$ and $(\epsp)_{\rm NP}$ from parameter 
scan using Eq.~\eqref{eq: scan}. The yellow points are 
obtained with no experimental constraint imposed, while blue 
points are obtained after imposing B sector constraints of Table~\ref{tab: B-meson-data}.
}
\label{fig: epsp-eps}
\end{figure}

In Fig.~\ref{fig: epsp-eps}, we show the scatter plot obtained
in the $\eps^{\rm NP}$ vs $(\epsp)_{\rm NP}$ plane. We 
purposely show parameter ranges that are far larger than 
allowed by data to illustrate the complementarity of the kaon 
sector as sensitive probes of $\rho_{ij}$ couplings. The blue 
(yellow) points are allowed (disallowed) by B physics data. For 
light $H^+$ (left), B sector basically rules out negative values 
of $\kappa$, but is less efficient in constraining positive 
$\kappa$ values. For $(\epsp)_{\rm NP}$, we notice an almost 
opposite situation: there is more population of scatter points 
corresponding to large and negative $(\epsp)_{\rm NP}$ values, 
while relatively few points for positive $(\epsp)_{\rm NP}$. 
{
For heavy $H^+$ case (right), the spread of the allowed points 
has shrunk because heavy $H^+$ suppresses NP contribution to 
$\kappa$ and $\epsp$, but now negative $\kappa$ values are also 
possible, because of relative inefficiency of B sector in 
constraining $\rho_{ct}$. We will expound on the significance 
of sizeable $\rho_{ct}$ later when we discuss $s\to d \nu \nu$ 
processes.} 

{
We note that constraints from B sector alone already restrict 
$\epsp$ values to the range $-5 \times 10^{-4} \lesssim 
(\epsp)_{\rm NP} \lesssim 2 \times 10^{-4}$ for light $H^+$, 
and $-1 \times 10^{-4} \lesssim (\epsp)_{\rm NP} \lesssim 5 
\times 10^{-4}$ for heavy $H^+$, consistent with current data 
(Eq.~\eqref{eq: epsp-bound}). {The maximum positive 
$(\epsp)_{\rm NP}$ is $\sim 3 \times 10^{-4}$ in heavy $H^+$ 
case after one imposes the $|\kappa| < 0.2$} constraint from 
$\eps$ data (Eq.~\eqref{eq: epsK-bound}). On the other hand, 
for light $H^+$ case, we note that although positive 
$(\epsp)_{\rm NP}$ of order ${\cal O}(10^{-4})$ can be reached, 
larger negative contributions are far more preferred, 
regardless of $\eps$ constraint. Figs.~\ref{fig: rct-rtt-B} 
and \ref{fig: epsp-eps} underline the point that $\eps$ 
provides complementary --- much better in many cases --- 
constraints on $\rho_{ij}$, and therefore must be included in 
any phenomenological study of g2HDM. In the following parameter 
scans, we impose the $\eps^{\rm NP}$ constraint from 
Eqs.~\eqref{eq: epsK-bound} along with $B$ physics constraints 
of Table~1.}

%

\subsection{ 
{The normalized ratios ${\cal R}_{\nu}^{+,0},\,
      {\cal R}_{\mu}^{L,S}$}}

Turning to rare kaon decays, we define four SM-normalized ratios:
\begin{align}
  {\cal R}_{\nu}^{+}
  = \frac{{\cal B}(K^+ \to \pi^+ \nu\bar \nu)}
       {{\cal B}(K^+ \to \pi^+ \nu\bar \nu)_{\rm SM}},
  ~
  {\cal R}_{\nu}^{0}
  = \frac{{\cal B}(K_L \to \pi^0 \nu\bar \nu)}
       {{\cal B}(K_L \to \pi^0 \nu\bar \nu)_{\rm SM}},
  ~
  {\cal R}_{\mu}^{L(S)}
  = \frac{{\cal B}(K_{L(S)} \to \mu\mu)}
         {{\cal B}(K_{L(S)} \to \mu\mu)_{\rm SM}},
\end{align}
which all become unity in SM-limit. We note that 
the g2HDM contribution to all rare kaon decays (as 
well as $B_q \to \mu^+\mu^-,\,B_d \to X_s \gamma$)
considered originate from penguin diagrams involving 
$H^+$ and a top or charm quark.\footnote{
{We tacitly drop the $u$-quark contribution from 
discussion. Flavor hierarchies suggest $|\rho_{tu}|, 
\, |\rho_{ut}|$ should be much less than charm 
counterparts, but we know rather little by direct 
measurement~\cite{Hou:2020ciy}.}
}
Therefore, the underlying flavor structure of NP 
contributions to these observables are similar if 
not highly correlated. We first give results for 
kaon decays as functions of $\kappa$ and 
$(\epsp)_{\rm NP}$. We subsequently highlight 
various correlations between these observables.

\begin{figure}[t]
\center
\includegraphics[width=6.7cm, height=5.5cm]{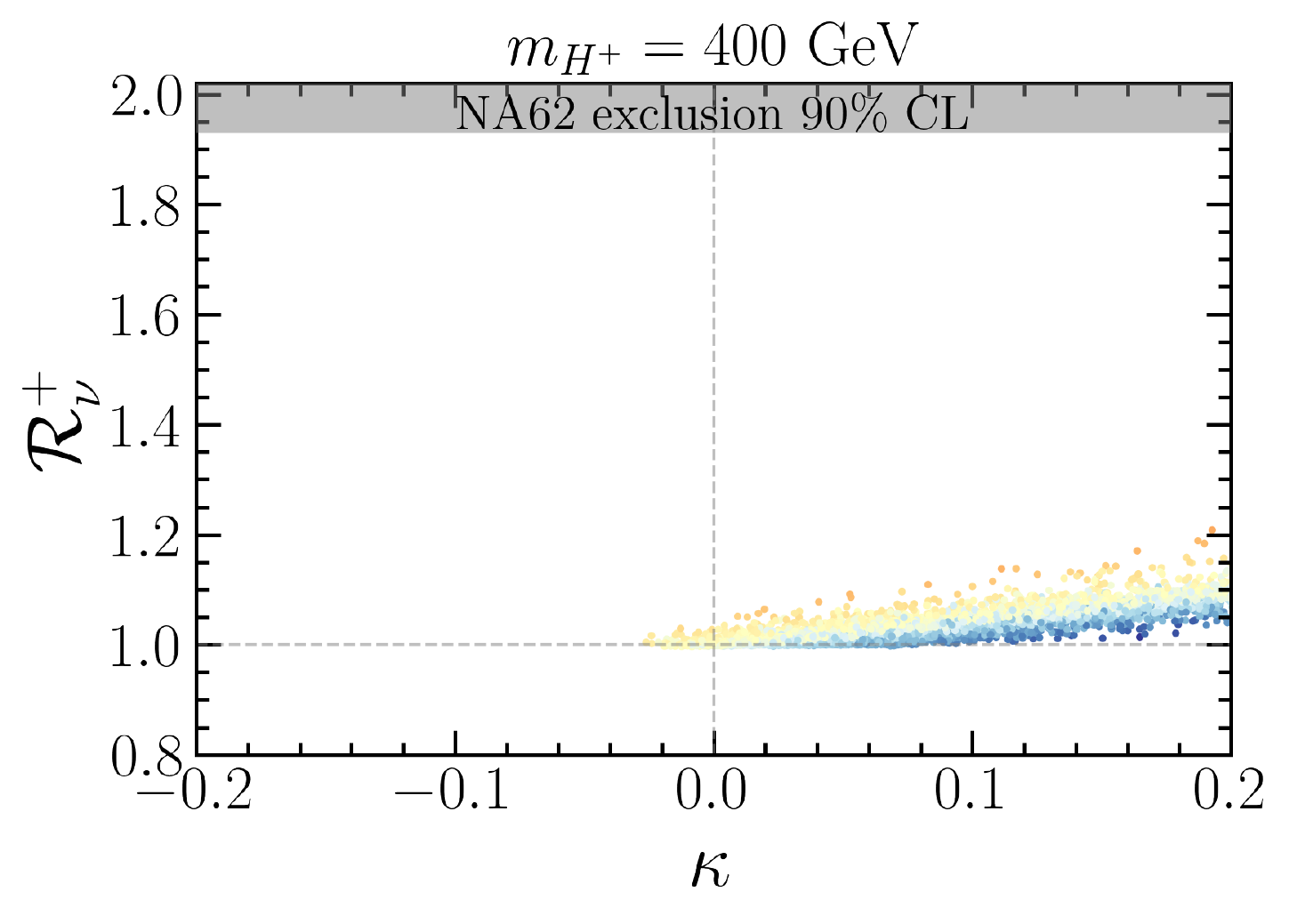}
\includegraphics[width=8.15cm, height=5.5cm]{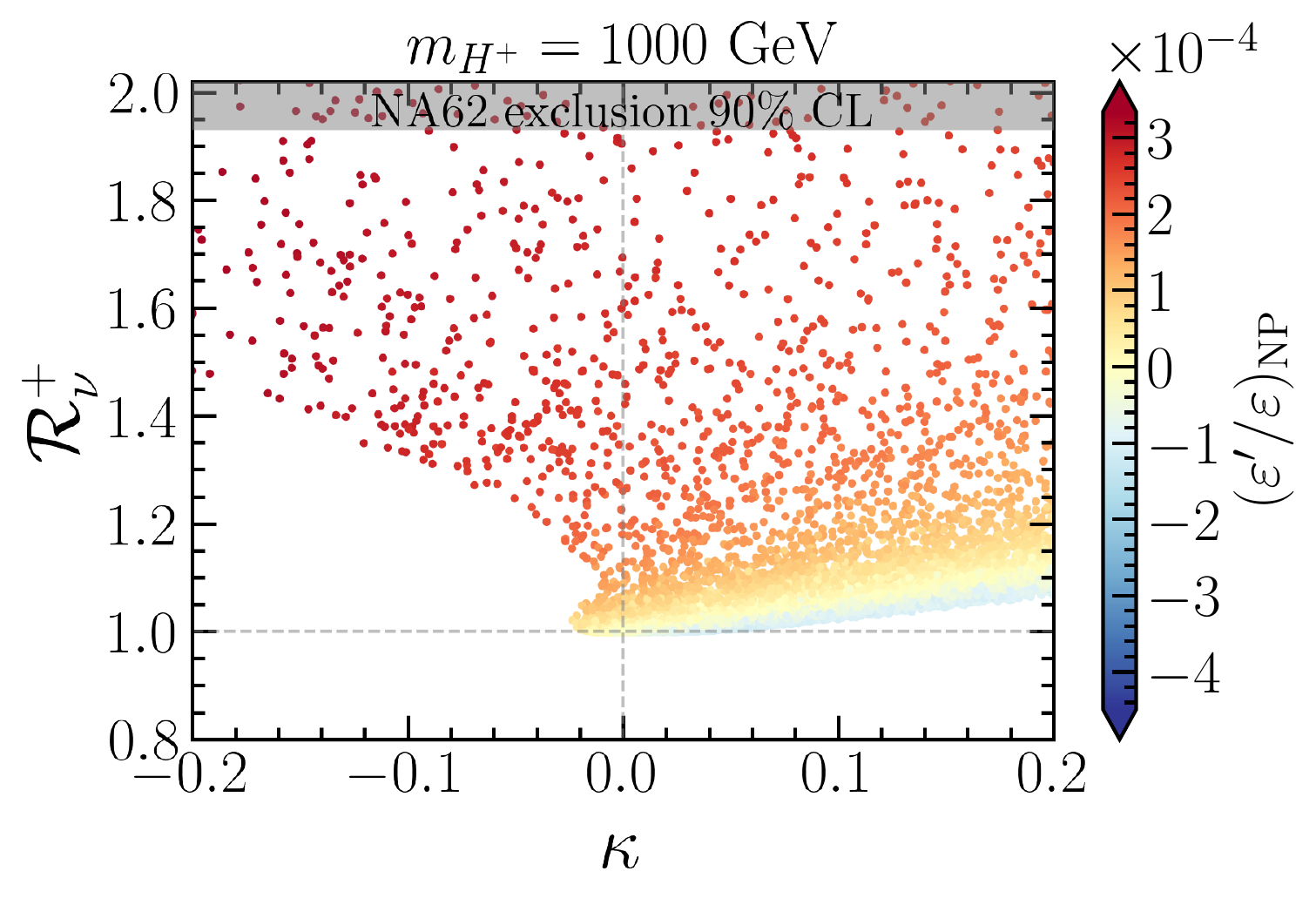}\\
\includegraphics[width=6.7cm, height=5.5cm]{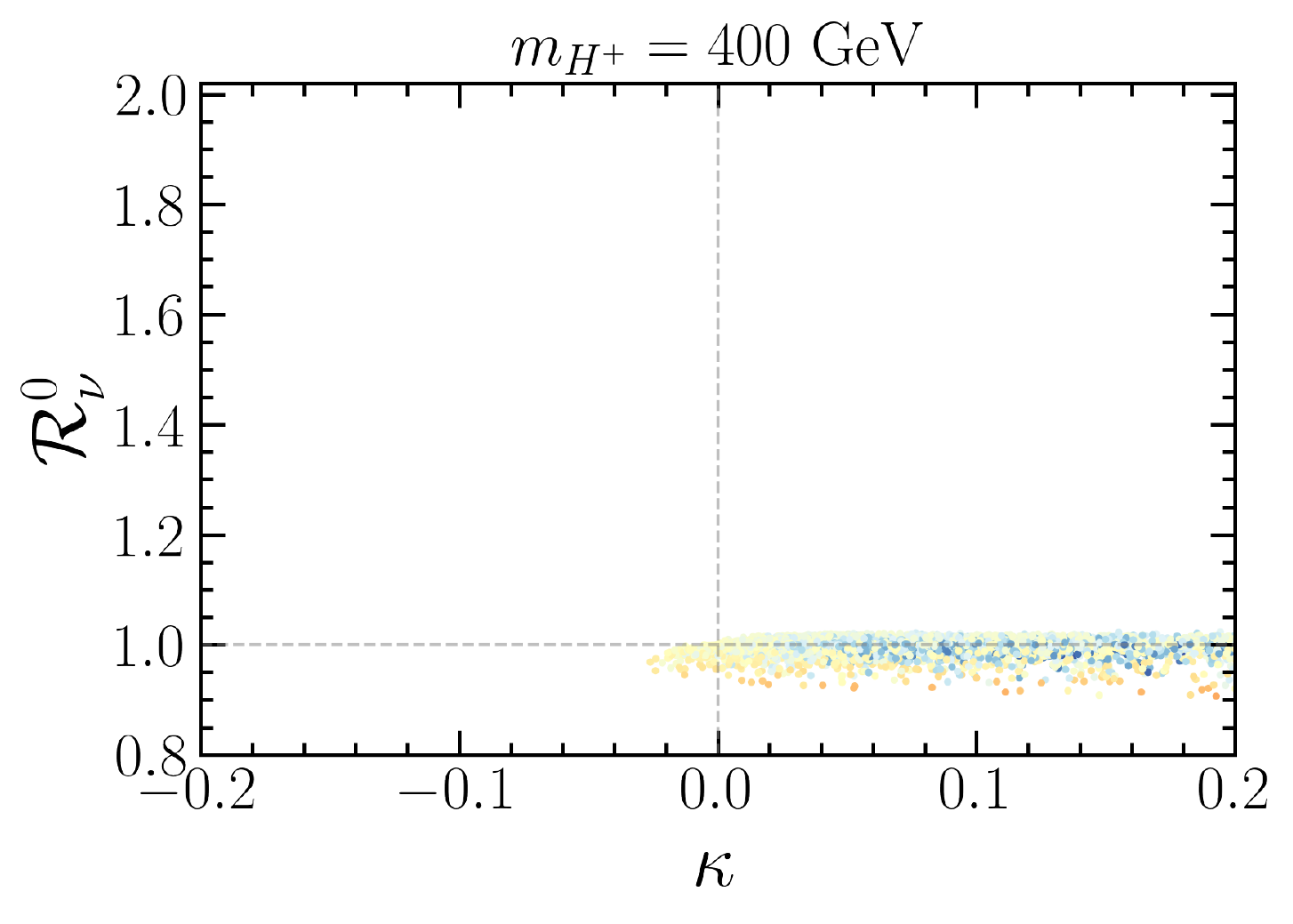}
\includegraphics[width=8.15cm, height=5.5cm]{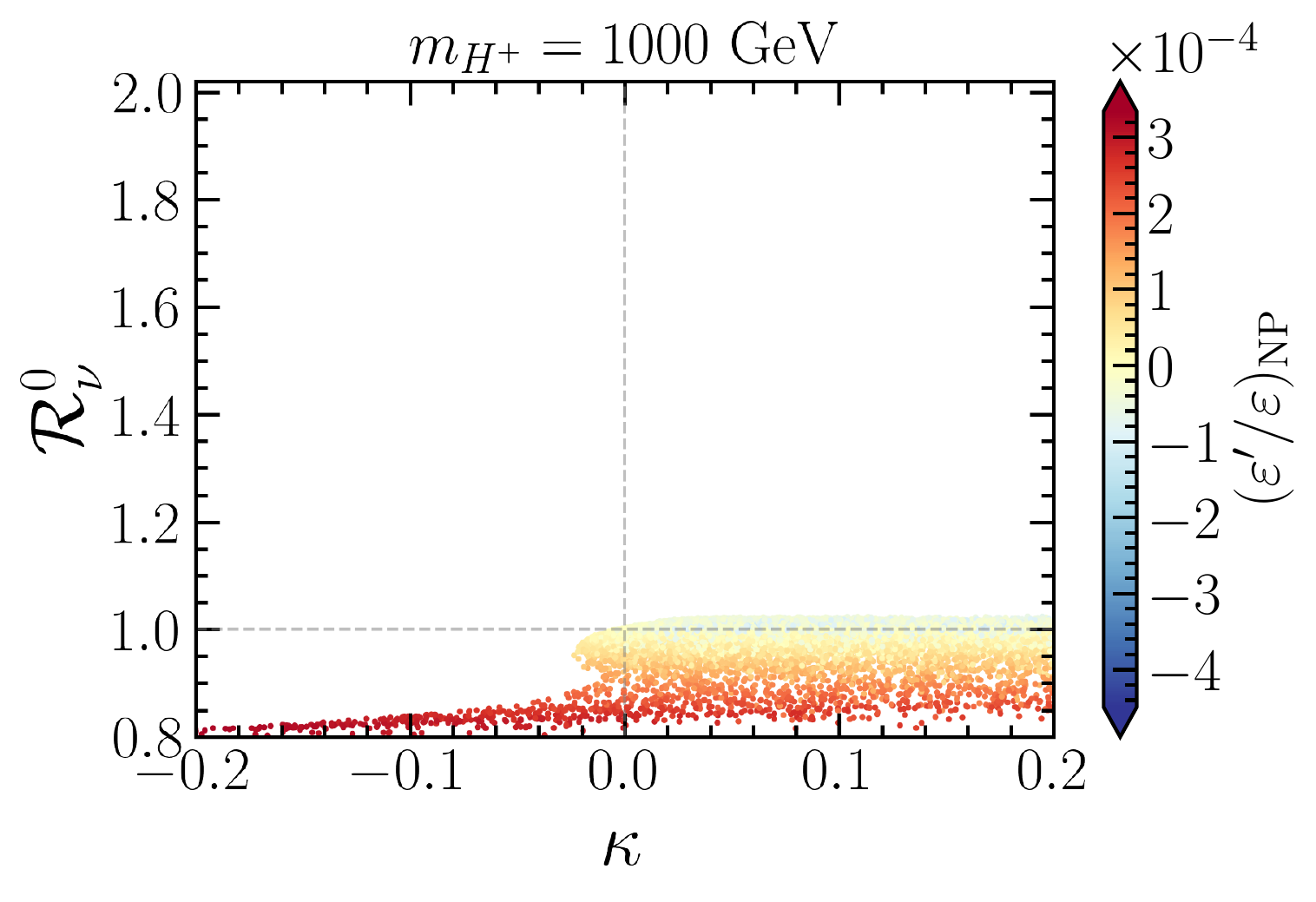}
\caption{
Ratios $\R_\nu^{+}$ (upper) and $\R_\nu^{0}$ (lower) 
as function of $\kappa$, and correlation with NP in $\epsp$.
}
\label{fig: epsK-Knn}
\end{figure}

We give scan values of $\R_\nu^+$ (upper) and 
$\R_\nu^0$ (lower) vs $\kappa$, i.e. $\eps^{\rm NP}$, 
in Fig.~\ref{fig: epsK-Knn}. For light $H^+$ (left), we 
find ${\cal B}(\Kpnn)$ can be enhanced to $10$--$20\%$ 
above SM. For $\KLnn$, somewhat opposite effect is 
noticed: the $\R_\nu^0$ remains either close to SM 
value or is {slightly suppressed; the largest 
suppression reaching about $10\%$. For heavy $H^+$ 
case (right), the results may appear a bit 
counterintuitive, as we see rather large effect for 
$\Kpnn$, which can be enhanced to the upper limit of 
NA62 (actually reaching up to $\R_\nu^+ \lesssim 4$), 
while $\KLnn$ can be suppressed by $20\%$ compared 
to SM value.}

We also illustrate in Fig.~\ref{fig: epsK-Knn} the 
correlation of rare $K$ decays with $(\epsp)_{\rm NP}$. 
Large positive $(\epsp)_{\rm NP}$ (red) correlates 
with larger effects in branching ratios, and negative  
values (blue) with smaller effects. Note also that for 
heavy $H^+$, one can still have large effects in 
$\R_\nu^{+, 0}$ despite $\kappa\sim 0$, especially for 
charged mode. {\it This highlights the importance 
of $\Kpnn$ decay as a sensitive probe of heavy $H^+$.}

\subsection{ 
{$K \to \pi\nu\nu$ sensitivity to heavy charged Higgs}}

Before proceeding further, let us understand why the 
heavy $H^+$ case shows surprisingly large {effects} 
compared with light $H^+$. The g2HDM contribution to 
$s\to d \nu\bar\nu$ is the second term of Eq.~\eqref{eq: 
XNP}. Expanding the g2HDM part explicitly in terms of 
$\rho_{ij}$ and CKM elements {according to Eq.~\eqref{eq: 
b2snunu-WC}}, focusing on the dominant top loop diagram 
we obtain {(dropping factor of $-\delta_{ab}/16\pi^2$)}
%
\begin{align}
   \frac{C_{LL}^{a, b}}{{v_t}} \;\ \Rightarrow\;\
 & \frac{{\sum_j(V^*_{js}\r_{jt})\sum_k(\r^*_{kt}V_{kd})}}
      {V_{ts}^* V_{td}}\; G_Z({m_t^2/m_{H^+}^2)} \nn \\
 & = \left(\rtt   + \frac{V_{cs}^*}{V_{ts}^*}\rct\right)
   \left(\rtt^* + \frac{V_{cd}}{V_{td}}\rct^*\right)
  G_Z({m_t^2/m_{H^+}^2)},
\label{eq: CLL-ckm}
\end{align}
where the two CKM factors associated with $\rct$,
$V_{cs}^\ast/V_{ts}^\ast\simeq -23.5 - 0.46\, i$, and
$V_{cd}/V_{td}\simeq-22.8 - 9.4\, i$, respectively, are quite 
sizable. Thus, the $s\to d \nu\bar\nu$ process has rather 
special sensitivity to $\rct$. Revisiting Fig.~\ref{fig: 
rct-rtt-B}, we recall that the combined constraints from B 
sector plus $\eps$ restrict $\rct$ to very small values for 
light $H^+$. But for heavy $H^+$ case, B sector constraints 
weaken considerably, then $\eps$ allows $\rct$ to become 
appreciable. 

\begin{figure}[ht]
\includegraphics[width=7.6cm,height=5.5cm]{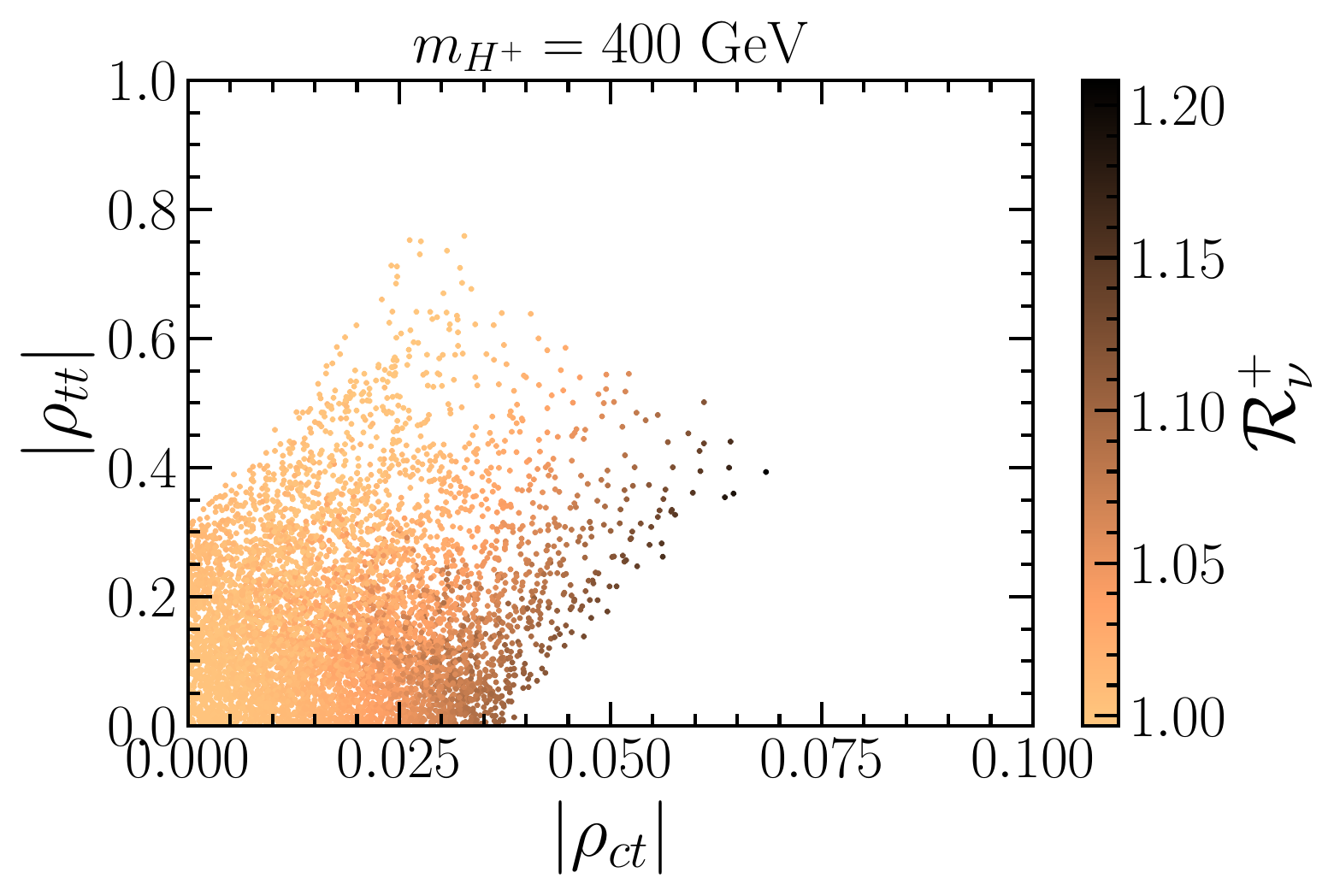}
\includegraphics[width=7.45cm,height=5.5cm]{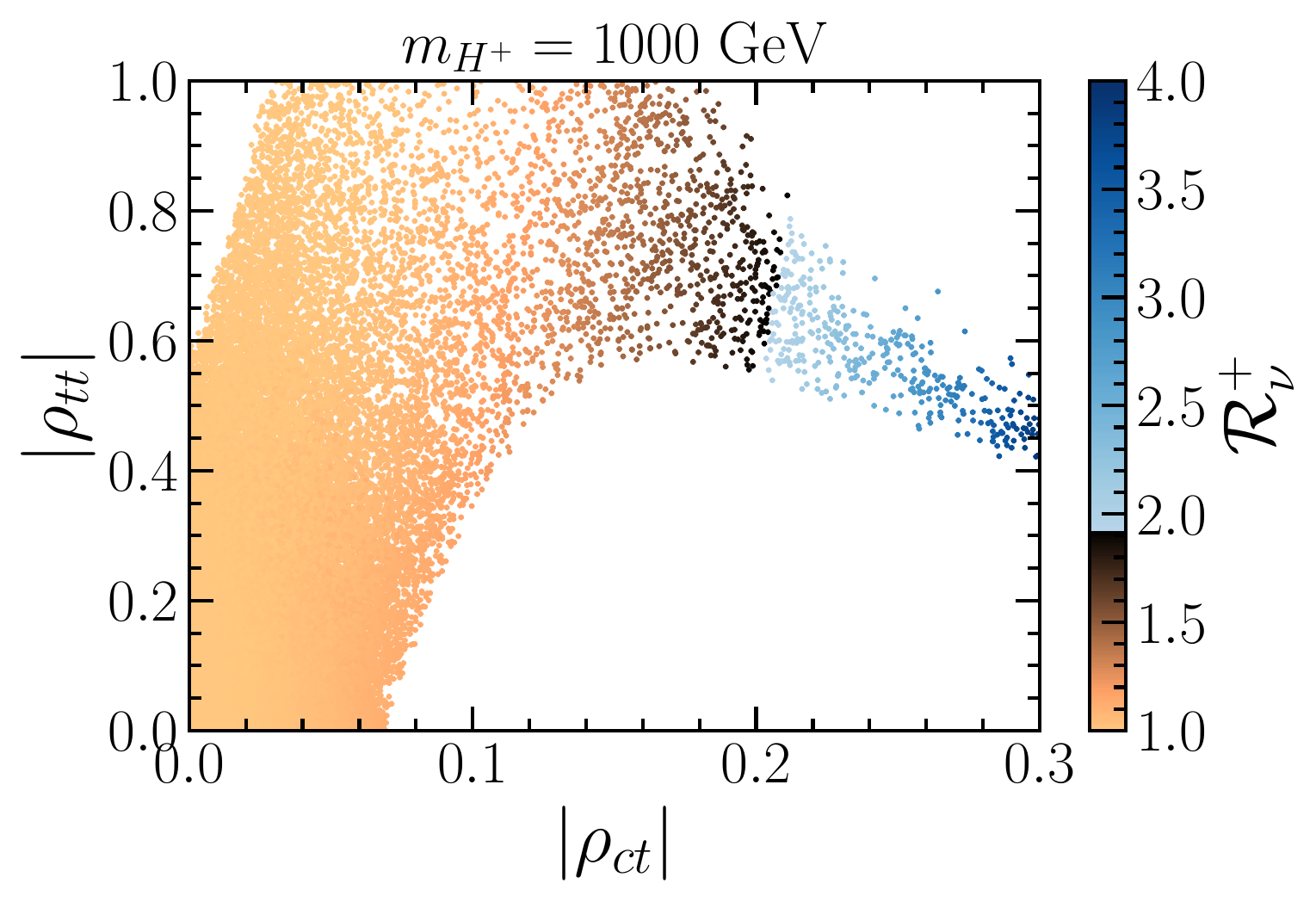}
\caption{
Scatter plot in $|\rct|$--$|\rtt|$, correlated with 
$\R_\nu^{+}$ after imposing flavor constraints.
}
\label{fig: rct-rtt-Kpnn}
\end{figure}

We show in Fig.~\ref{fig: rct-rtt-Kpnn} the scatter in the
$|\rct|$--$|\rtt|$ plane, in correlation with $\R_\nu^{+}$. 
For light $H^+$, $|\rct|$ is highly constrained below 
$\sim 0.06$, resulting in marginal enhancement of 
$\R_\nu^{+}$. But for heavy $H^+$, much larger $\rct$ up 
to $\sim 0.2$ is allowed, and the aforementioned CKM 
factors then boost $\R_\nu^{+}$ up to the NA62 limit. 
In fact, flavor constraints (B sector together with $\eps$) 
{do not rule out higher values (illustrated in blue) of 
$\R_\nu^{+}$, but cut off only by the NA62 bound itself. 
This further illustrate the significant role of $\Kpnn$ 
in probing heavy $H^+$, which was} already alluded to 
in the previous subsection.  Note also that enhancing 
$\R_\nu^{+}$ requires a significant value for $\rtt$, 
but still less than unity. The dark points that 
saturate the NA62 bound in  Fig.~\ref{fig:  rct-rtt-Kpnn}(right) are for $|\rct|$ reaching 
$\sim 0.2$, which one would not have anticipated 
based on the behavior seen in Fig.~\ref{fig: 
rct-rtt-Kpnn}(left) for lighter $m_{H^+}$.

The contrasting behavior of the CP-violating counterpart 
$\KLnn$ can be understood by noting that the purely 
$\rho_{ct}$ term, which enjoys the largest CKM 
enhancement $|v_c/v_t| \simeq 580 $, belongs to the 
CP-conserving part of NP, so $\KLnn$ is not sensitive 
to this term. But $\rtt$-$\rct$ interference terms can 
have CP-violating phase. {After expanding 
Eq.~\eqref{eq: CLL-ckm} in terms of $\rho_{ij}$ and CKM  elements, then putting back in Eq.~\eqref{eq: XNP}, 
one obtains,
\begin{align}
	v_t X_{\rm eff}
 \simeq (-5.0 + 2.2 i ) &- (96.4 + 0.06i) |\rho_{ct}|^2
	- (0.15 - 0.06i) |\rho_{tt}|^2 \nn\\
	&+~ (4.1 - 0.08i) \rho_{tt}\rho_{ct}^\ast
	+ (3.6 - 1.5i) \rho_{tt}^\ast \rho_{ct},
\label{eq: sdnunu-expansion}
\end{align}
where the first term is the SM contribution. We first 
note that the $|\rho_{ct}|^2$ and $|\rho_{tt}|^2$ terms 
are close to real, but $\rho_{tt}^\ast \rho_{ct}$ is
complex. The imaginary part of $\rho_{tt}^\ast \rho_{ct}$ 
largely cancel between the two interference terms  
(and further suppressed by $|\rct| < 0.2$). For the 
real part, the $-0.08i$ coefficient to $\rho_{tt}^\ast 
\rho_{ct}$ is small, but carries the same sign as the 
$-1.5i$ coefficient to $\rho_{tt} \rho_{ct}^\ast$, 
which explains the destructive interference with SM 
effect, as seen in Fig.~\ref{fig: epsK-Knn}. 
}

Note that compared with $\rct$, there is no similar 
sensitivity to $\rtc$. In fact, flavor constraints on 
$\rtc$ are the poorest among the three top 
$\rho_{ij}$ couplings considered. This is due to two 
reasons: first, $\rtc$ is associated with the charm 
loop rather than top, hence the loop function is 
small; second, there is no CKM enhancement, i.e. 
$C_{LL}^{a, b}/v_t \propto |\rtc|^2$.

We have commented that the well-measured $\KLmm$ 
seems to {prefer LD $A_{L\g\g}$ effect to be
destructive against SD} in Eq.~\eqref{eq: KLmm-exp}. 
But $K_{L,S} \to \mu^+\mu^-$ interference can probe 
the sign of $A_{L\g\g}$~\cite{DAmbrosio:2017klp}. 
If constructive SD-LD interference turns out to be 
favored by data in the future, we see from 
Eq.~\eqref{eq: KLmm-SM} that the SM value is 
considerably higher than the experimental result of 
Eq.~\eqref{eq: KLmm-exp}. But $\KLmm$ behaves 
similarly to $\Kpnn$, since one just replaces 
$Z\nu\nu$ by $Z\mu\mu$ in the diagrams, hence g2HDM 
effect always {\it enhance} the branching ratio (see 
Fig.~\ref{fig: epsK-Knn}) hence can never match 
Eq.~\eqref{eq: KLmm-exp}. Thus, {g2HDM in the
parameter space we consider} cannot offer a solution 
to the potential new emergent tension. For $\KSmm$, 
we find variation in $\R_\mu^S$ is never more than 
$2\%$, which we doubt LHCb can distinguish. Thus, 
neither $\KLmm$ nor $\KSmm$ are interesting in g2HDM.

CKM enhancement factors analogous to Eq.~\eqref{eq: 
CLL-ckm} was first touched upon in the discussion of 
$b \to s\gamma$ transitions~\cite{Altunkaynak:2015twa}, 
where only one CKM factor of $\rct$ gets 
$1/\lambda^2$-enhanced ($\lambda = |V_{us}|$), the 
other being $\lambda^2$-suppressed. Among the 
{three ($b \to s, d$ and $s \to d$) type of 
penguins involving $H^+$-top quark in the loop, 
the $s \to d$ penguin is unique in receiving double 
$1/\lambda^2$-enhancement for both $\rct$ factors.} 
Analogous subtle CKM enhancement effects have been 
discussed for the tree level~\cite{Hou:2019uxa} 
$B \to \ell\nu$ process, now between $\bar ub$ and 
$\bar\ell\nu$ bilinears. It was stressed that, if 
Belle~II found the ratio ${\cal B}(B \to 
\mu\nu)/{\cal B}(B \to \tau\nu)$ would deviate from 
the SM value of 0.0045, it would not only rule out 
SM, but type~II 2HDM as well~\cite{Chang:2017wpl}, 
while proving $\rho_{tu} \neq 0$ in g2HDM. 
Another application of such CKM enhancement, rooted 
in the charged Higgs Yukawa interaction in 
Eq.~\ref{eq: Lag}, is the tree level $cg \to bH^+$ 
production process~\cite{Ghosh:2019exx} mentioned 
in the Introduction, which is surprisingly efficient 
compared with intuition derived from type~II 2HDM.

\subsection{ 
{Correlations in $K \to \pi\nu\nu$
   decays, and with $B_s \to \mu\mu$}}
%
\begin{figure}[t]
\includegraphics[width=6.7cm, height=5.5cm]{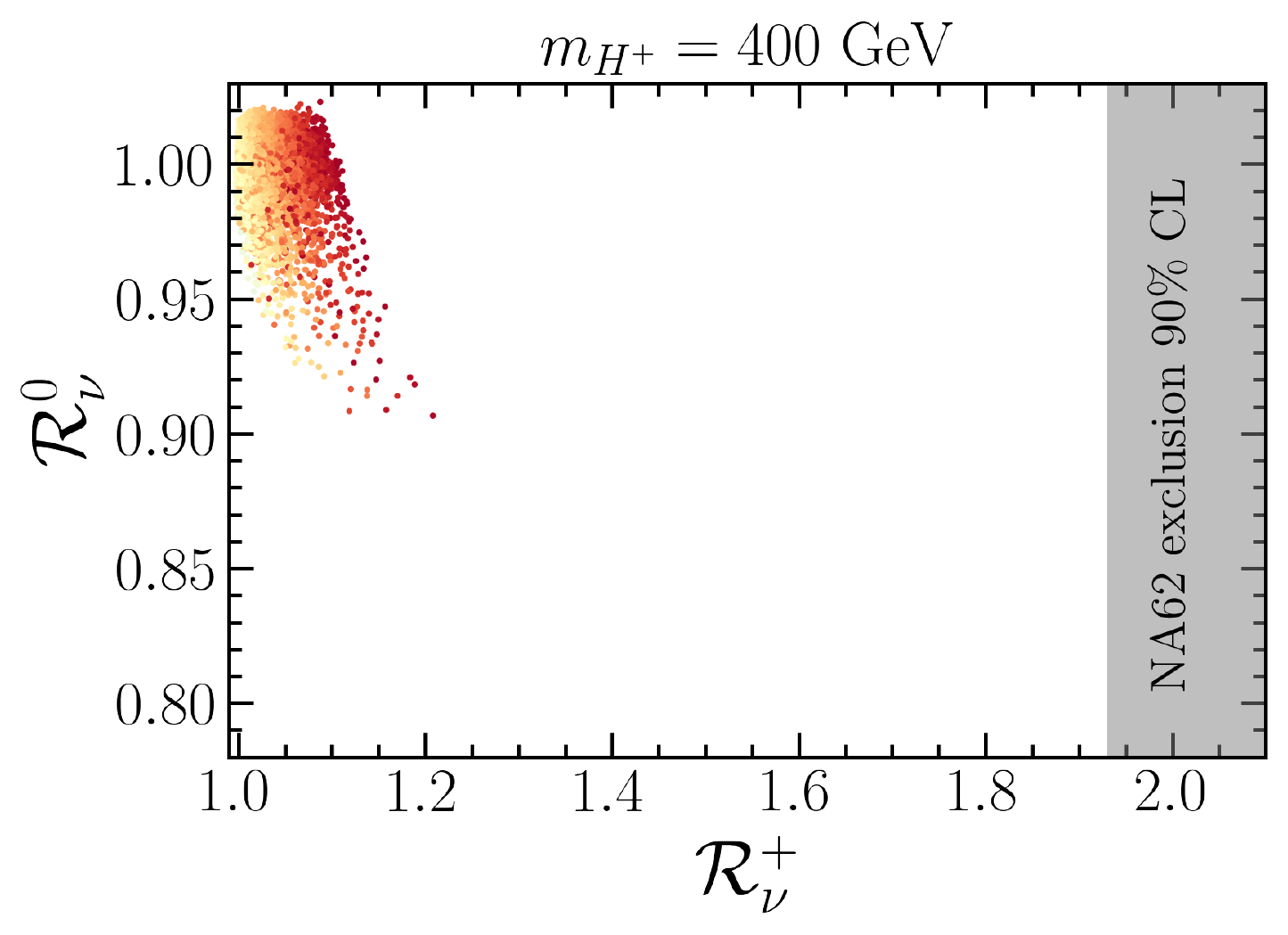}\includegraphics[width=8.9cm, height=5.5cm]{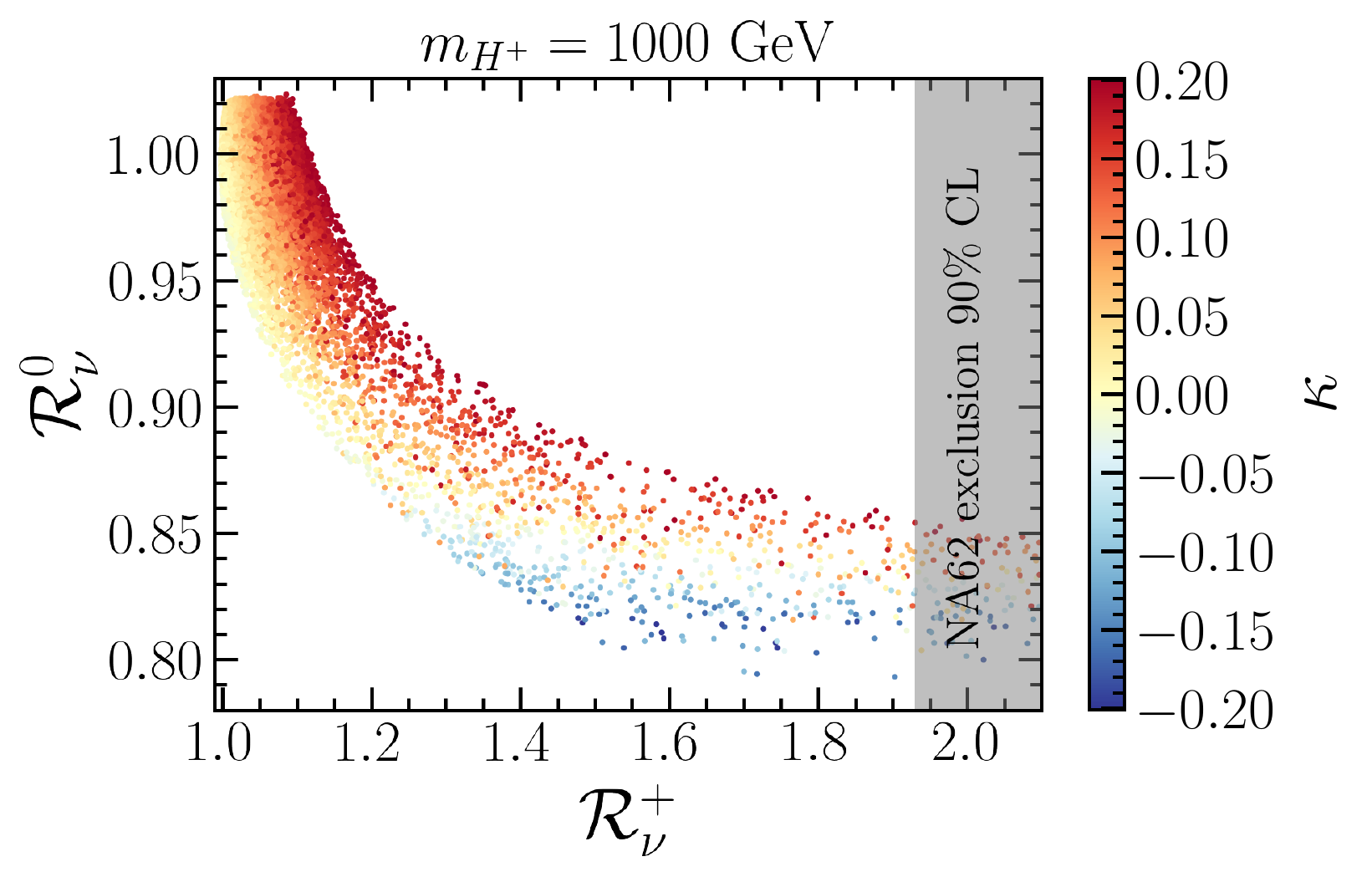}
\includegraphics[width=6.7cm, height=5.5cm]{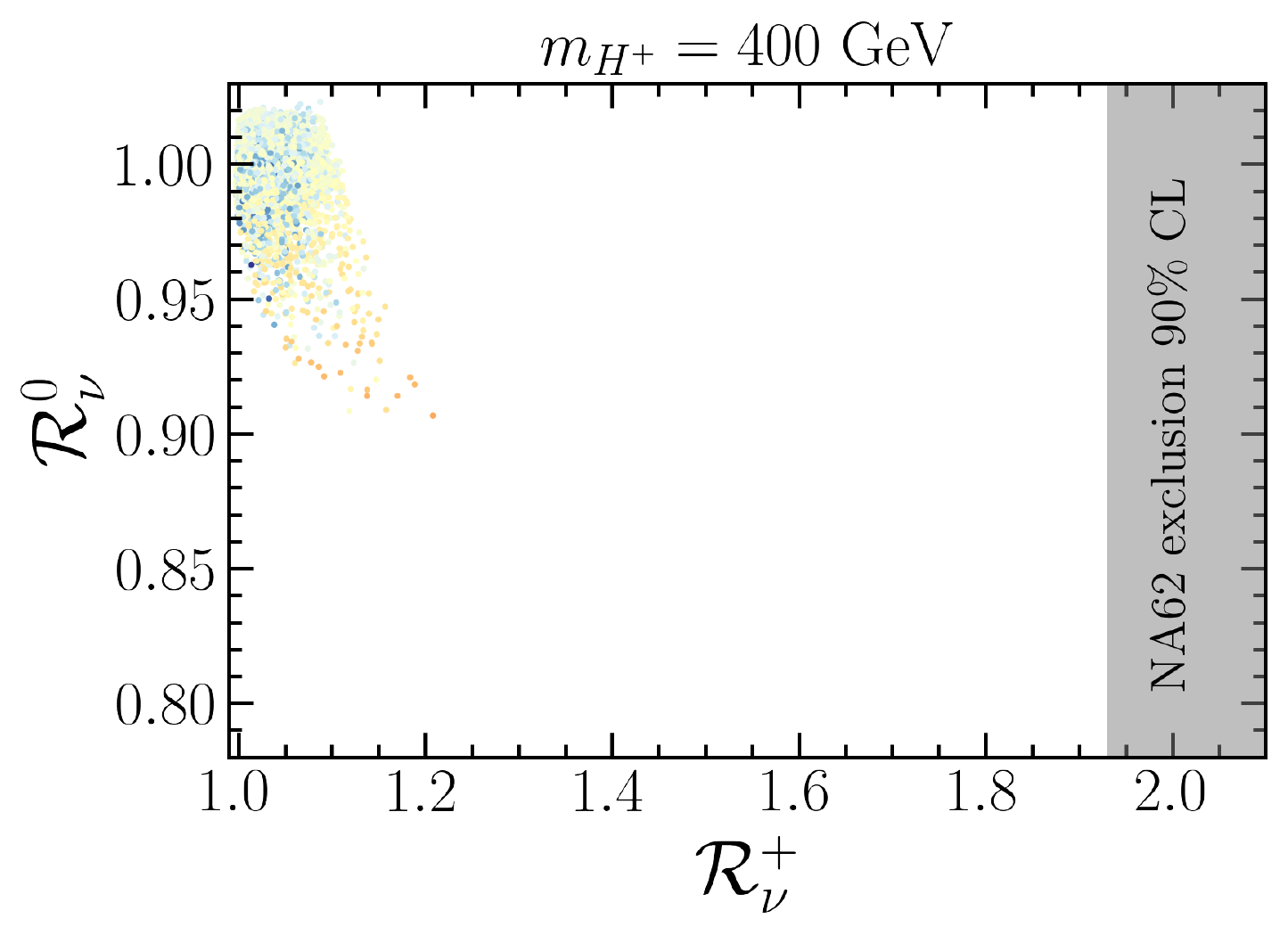}\includegraphics[width=8.4cm, height=5.5cm]{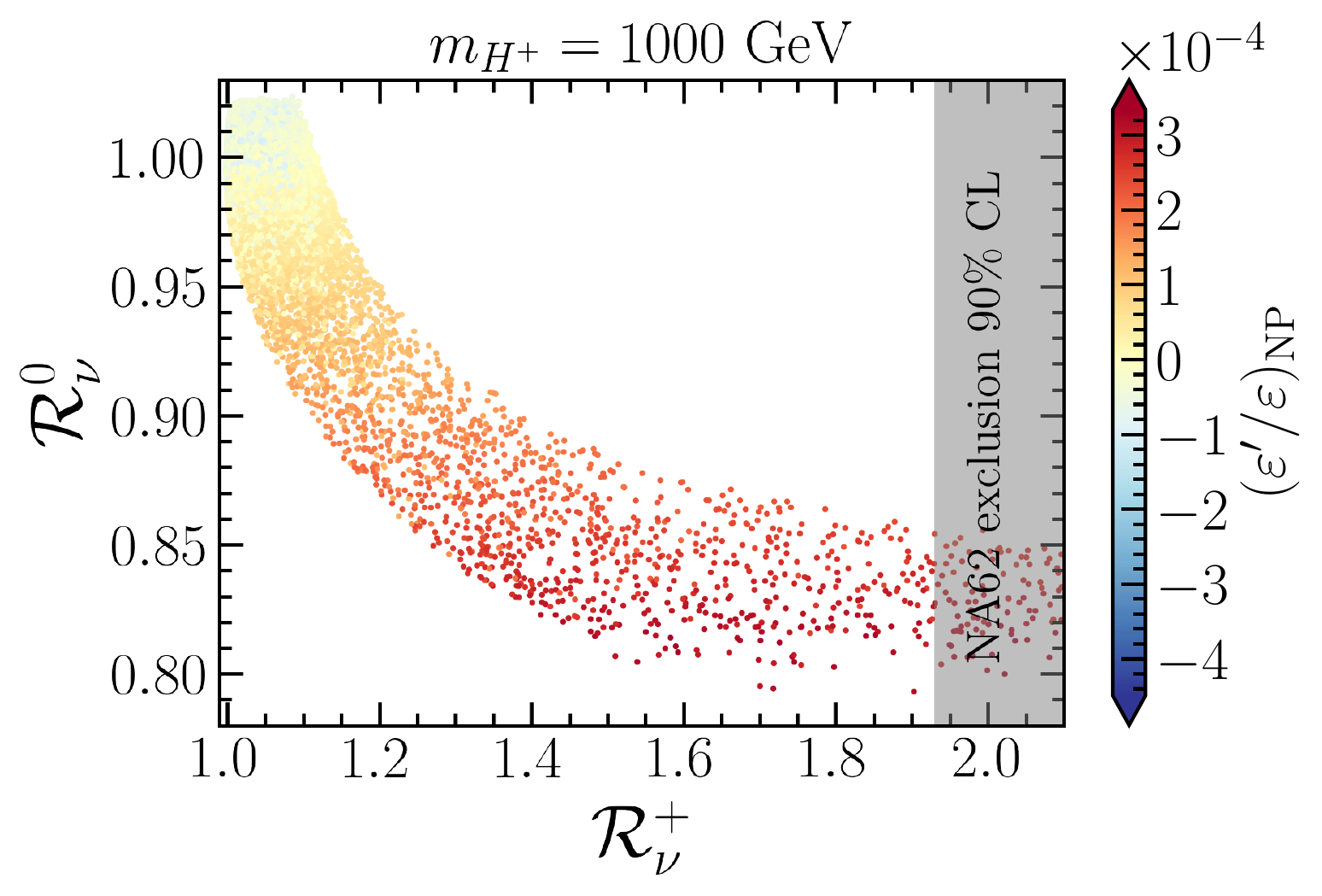}
\caption{
Correlations between $\Kpnn$, $\KLnn$,
and $\kappa$ (upper) and $(\epsp)_{\rm NP}$ (lower).
}
\label{fig: corr-K}
\end{figure}
We now discuss the correlations between rare $K$ decays, 
and implications for g2HDM. 

In Fig.~\ref{fig: corr-K}, we show the correlation of 
$\Kpnn$ and $\KLnn$ by plotting $\R_\nu^{+}$ vs 
$\R_\nu^{0}$ as functions of $\kappa$ (upper row) and 
$(\epsp)_{\rm NP}$ (lower row). As already noted from 
Fig.~\ref{fig: epsK-Knn}, while opposite in behavior, 
the g2HDM effects are far more pronounced in $\Kpnn$ 
compared with $\KLnn$, for both $m_{H^+} =$ 400 and 
1000 GeV. 
Needless to say, the decays conform with the 
Grossman-Nir bound, ${\cal B}(\KLnn) \lesssim 4.3\, 
{\cal B}(\Kpnn)$~\cite{Grossman:1997sk}. The future 
$10\%$ measurement of ${\cal B}(\Kpnn)$ by NA62, 
expected by 2024, {could start to} limit the amount 
of suppression possible for $K_L$ decay. The fact that 
large negative values (blue points) of $\kappa$ 
correlates with higher enhancement of $\Kpnn$ and 
suppression of $\KLnn$ was already obvious from 
Fig.~\ref{fig: epsK-Knn}, but it becomes visually 
more distinct in Fig.~\ref{fig: corr-K} (upper row), 
where blue points lie on the lower end of the scatter 
plot. The opposite correlation is seen for 
$(\epsp)_{\rm NP}$, where large positive values (red 
points) lie on the lower end of the scatter plot, while 
negative contributions (blue points) are closer to SM 
value. The upshot is that, with improved results from 
NA62 expected soon, we find $\Kpnn$ would play 
the leading role in probing g2HDM in the coming 
future. The $\KLnn$ mode {can play the crucial role} 
of confirming the unique g2HDM effects 
in the future, but the task would be challenging as 
it may require upgrades beyond KOTO 
Step-2~\cite{Nomura:2020oor, Aoki:2021cqa} and 
KLEVER~\cite{KLEVERProject:2019aks}. Similarly, 
measurement of $\epsp$ at sensitivity of $10^{-4}$ is 
needed to probe g2HDM effects, which at present looks 
unlikley. Once again, this makes $K^+\to\pi^+\nu\bar\nu$ 
the leading kaon observable to watch out for.

\begin{figure}[t]
\center
\includegraphics[width=6.7cm, height=5.5cm]{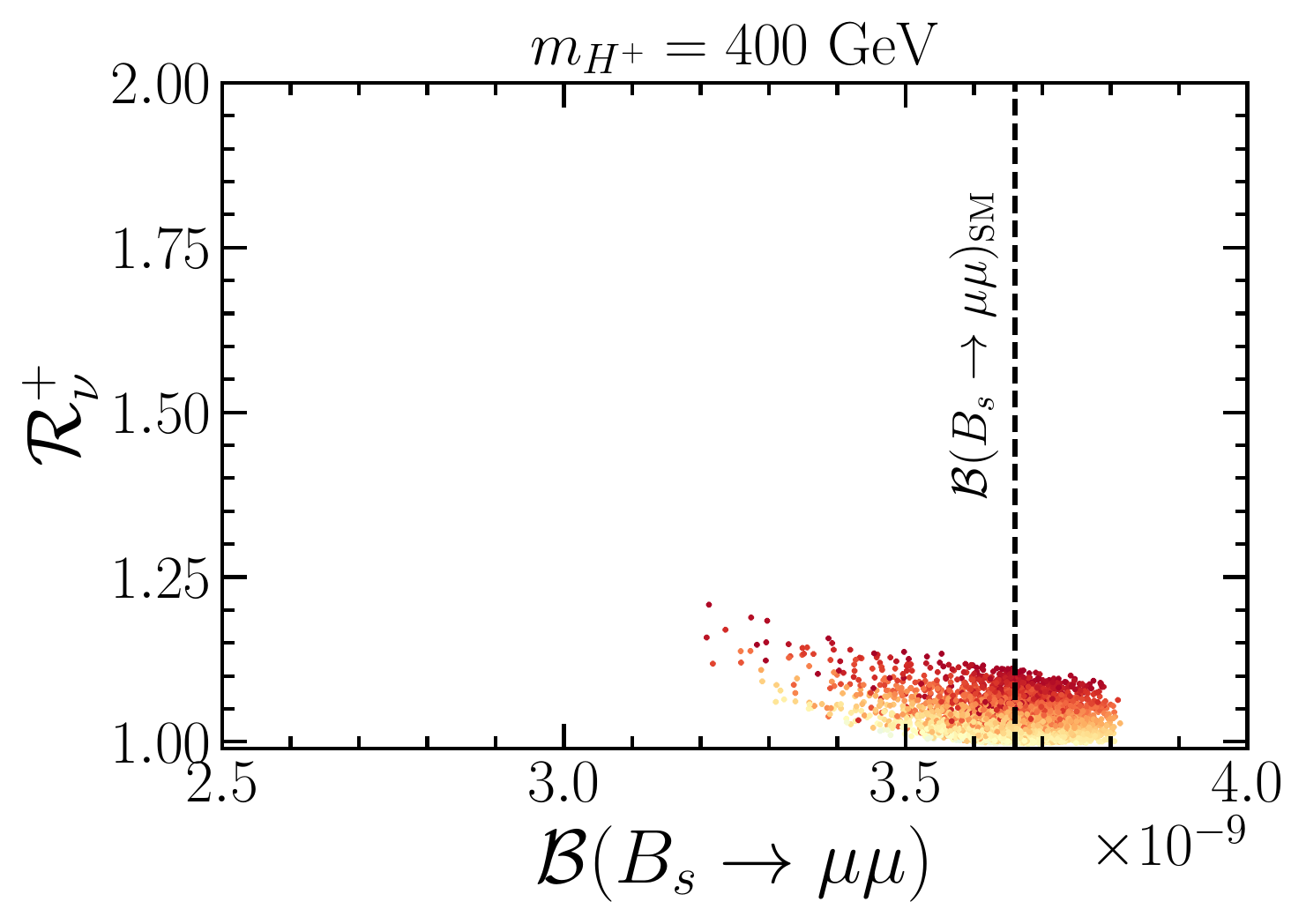}\includegraphics[width=8.9cm, height=5.5cm]{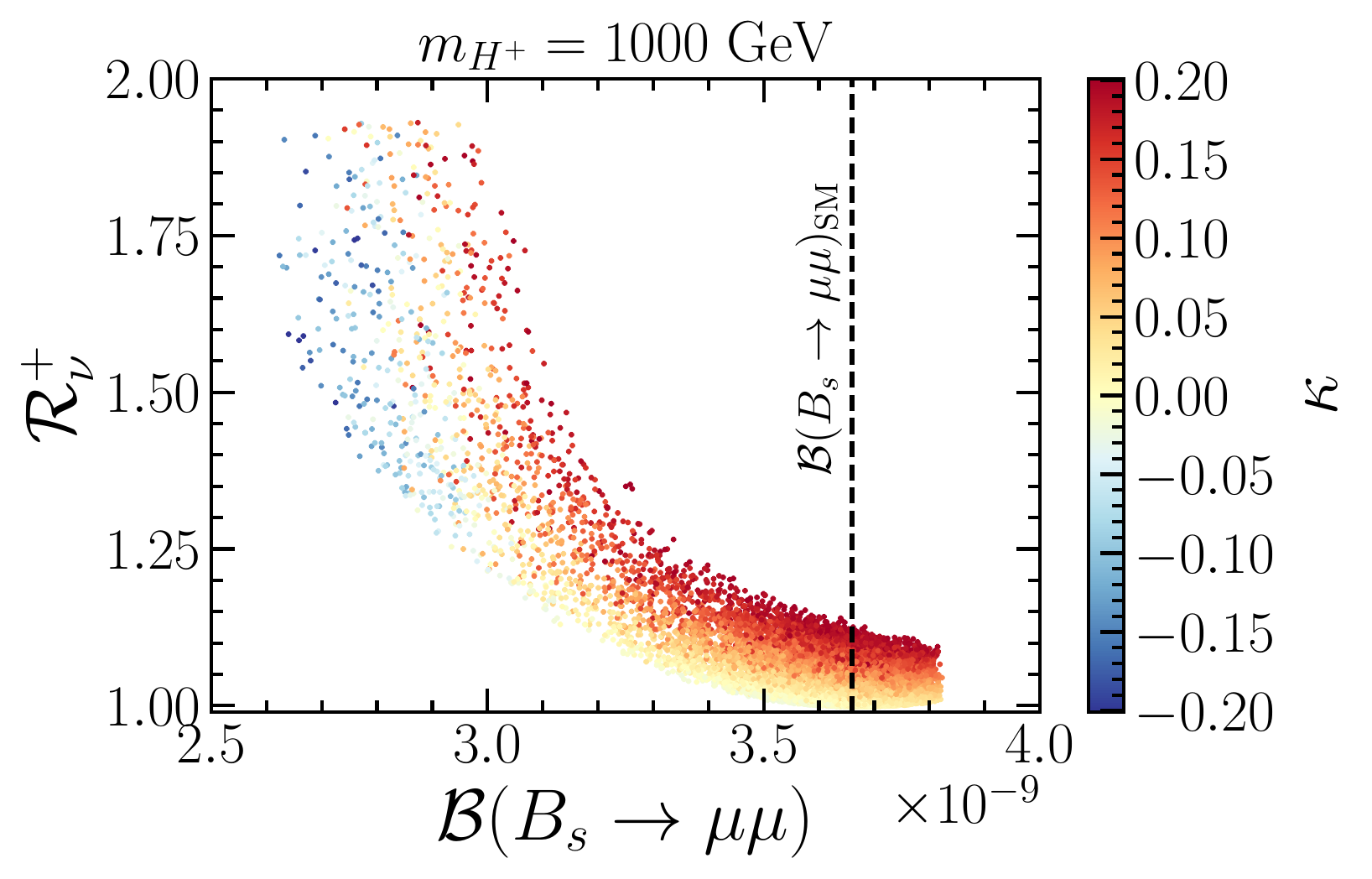}
\includegraphics[width=6.7cm, height=5.5cm]{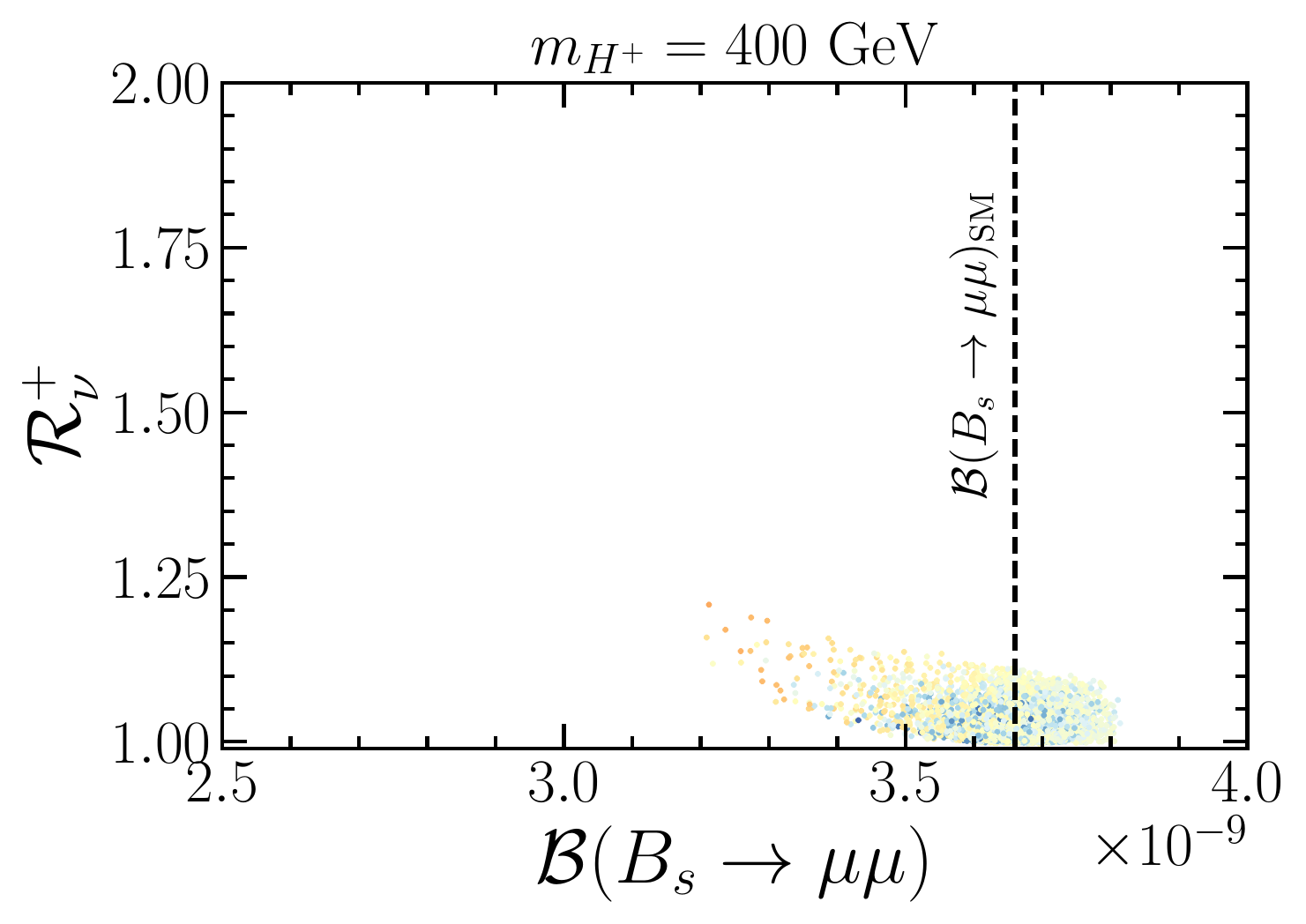}\includegraphics[width=8.4cm, height=5.5cm]{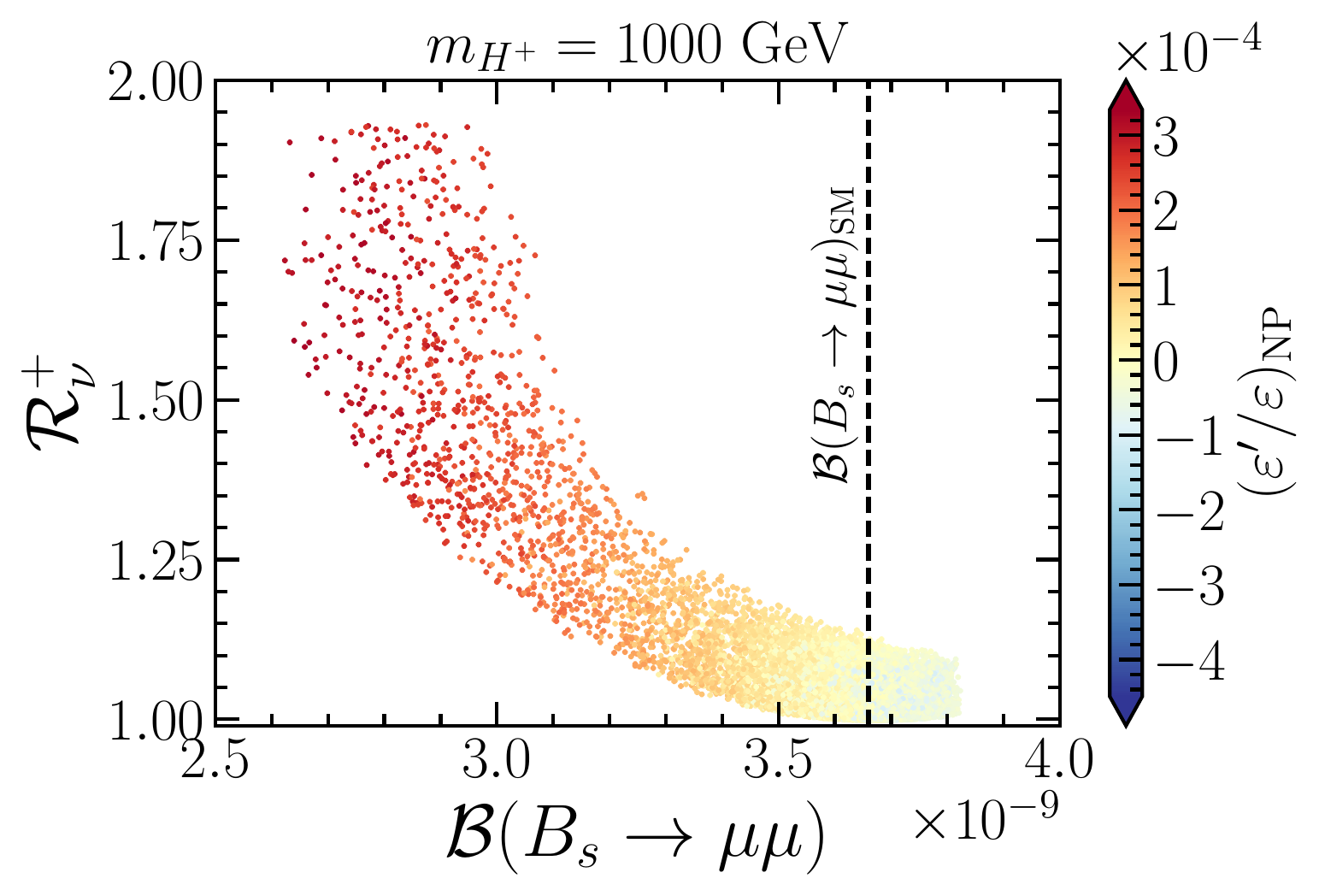}
\caption{
Correlations between $\Kpnn$, $B_s\to\mu^+\mu^-$,
and $\kappa$ (upper) and $(\epsp)_{\rm NP}$ (lower).}
\label{fig: kpnn-Bsmm}
\end{figure}

Finally, it is of considerable interest to discuss the 
correlation between $\Kpnn$ and $B_q \to \mu^+\mu^-$. 
In Fig.~\ref{fig: kpnn-Bsmm}, we plot ${\cal B}(B_s \to 
\mu^+\mu^-)$ vs $\R_\nu^{+}$, together with $\kappa$ 
(upper row) and $(\epsp)_{\rm NP}$ (lower row). For 
light $H^+$, ${\cal  B}(B_s \to \mu^+\mu^-)$ mostly 
stays within $2\sigma$ range of the SM value of $(3.66 
\pm 0.14) \times 10^{-9}$~\cite{Beneke:2019slt}, where
dashed line indicates the central value, while 
$\R_\nu^{+}$ changes by less than $\sim 20\%$; large 
suppression of $B_s\to\mu\mu$ coupled with enhancement 
of $\Kpnn$ is ruled out by current flavor data. But 
for heavy $H^+$, the anti-correlation of g2HDM effects 
in ${\cal B}(B_s \to \mu^+\mu^-)$ and $\R_\nu^{+}$ is 
clearly visible. This is of interest because the central 
value of LHCb~\cite{LHCb:2021vsc, LHCb:2021awg} at 
${\cal B}(B_s \to \mu^+\mu^-)$ at $3.09 \times 10^{-9}$ 
is somewhat lower than SM, but the new full Run 2 result 
reported by CMS is fully consistent with SM.\footnote{
The even smaller central value of ${\cal B}(B_s \to 
\mu^+\mu^-)_{\rm ave} \simeq 2.69 \times 10^{-9}$ from 
combined~\cite{LHCb:2020zud} ATLAS, CMS and LHCb analysis 
based on data collected during 2011-2016 probably should 
no longer be considered.
} 
While the earlier impression that $B_s\to\mu\mu$ 
is slightly below SM seems to have gone away, {some 
insight may be gained from kaon decay.} The right plot 
of Fig.~\ref{fig: kpnn-Bsmm} shows that ${\cal B}(B_s 
\to \mu^+\mu^-)$ can be suppressed in g2HDM for heavy 
$H^+$. The correlation with higher values of 
$\R_\nu^{+}$ then means the upcoming NA62 measurement 
could provide early tests of the size of g2HDM effects 
in $B_s\to\mu\mu$. This correlation is notable for both 
$\kappa$ and $(\epsp)_{\rm NP}$ in opposite way, 
{more prominent for negative values of $\kappa$, 
but positive values for $(\epsp)_{\rm NP}$.}
%
\begin{figure}[h]
\center
\includegraphics[width=6.7cm, height=5.5cm]{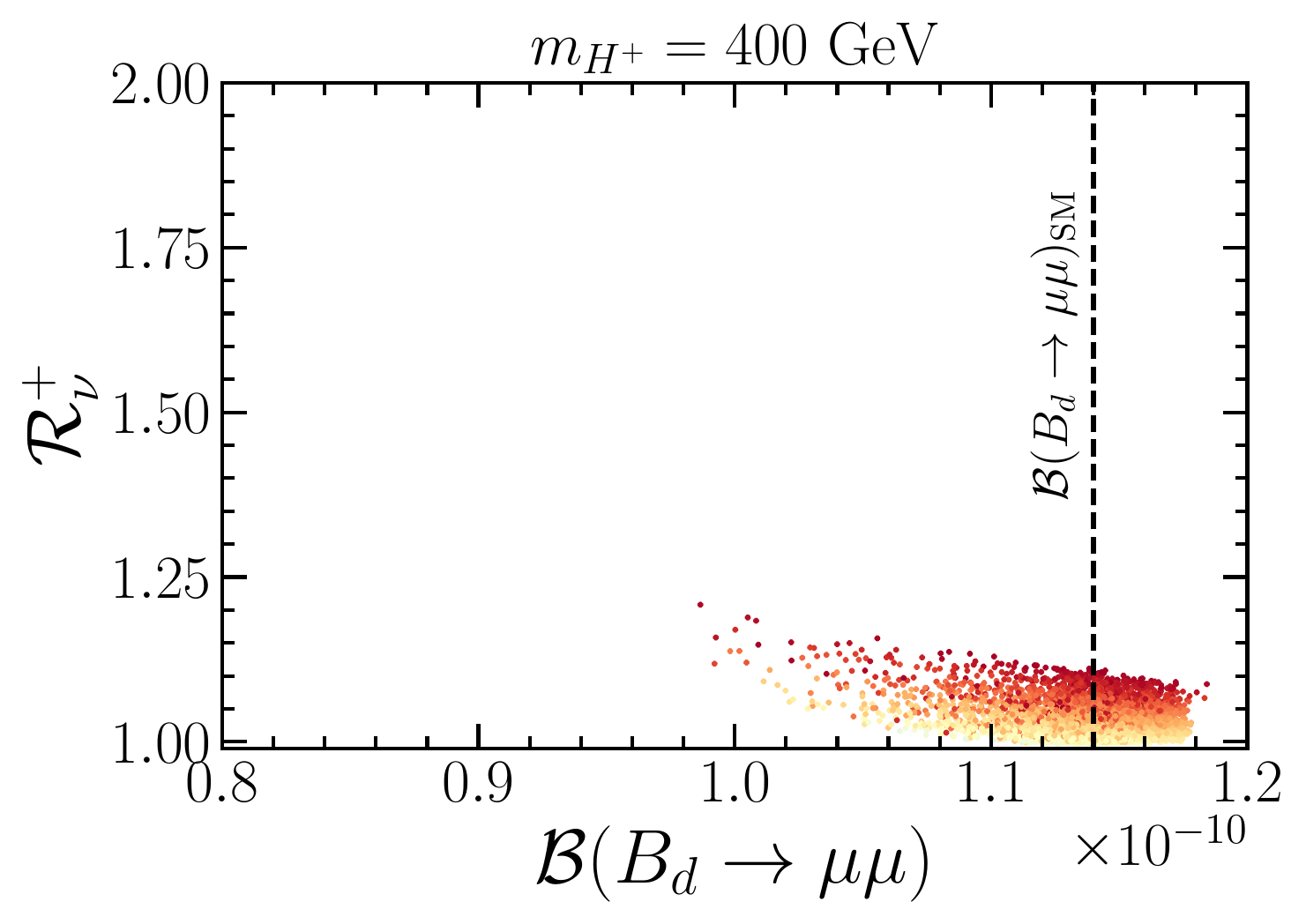}\includegraphics[width=8.9cm, height=5.5cm]{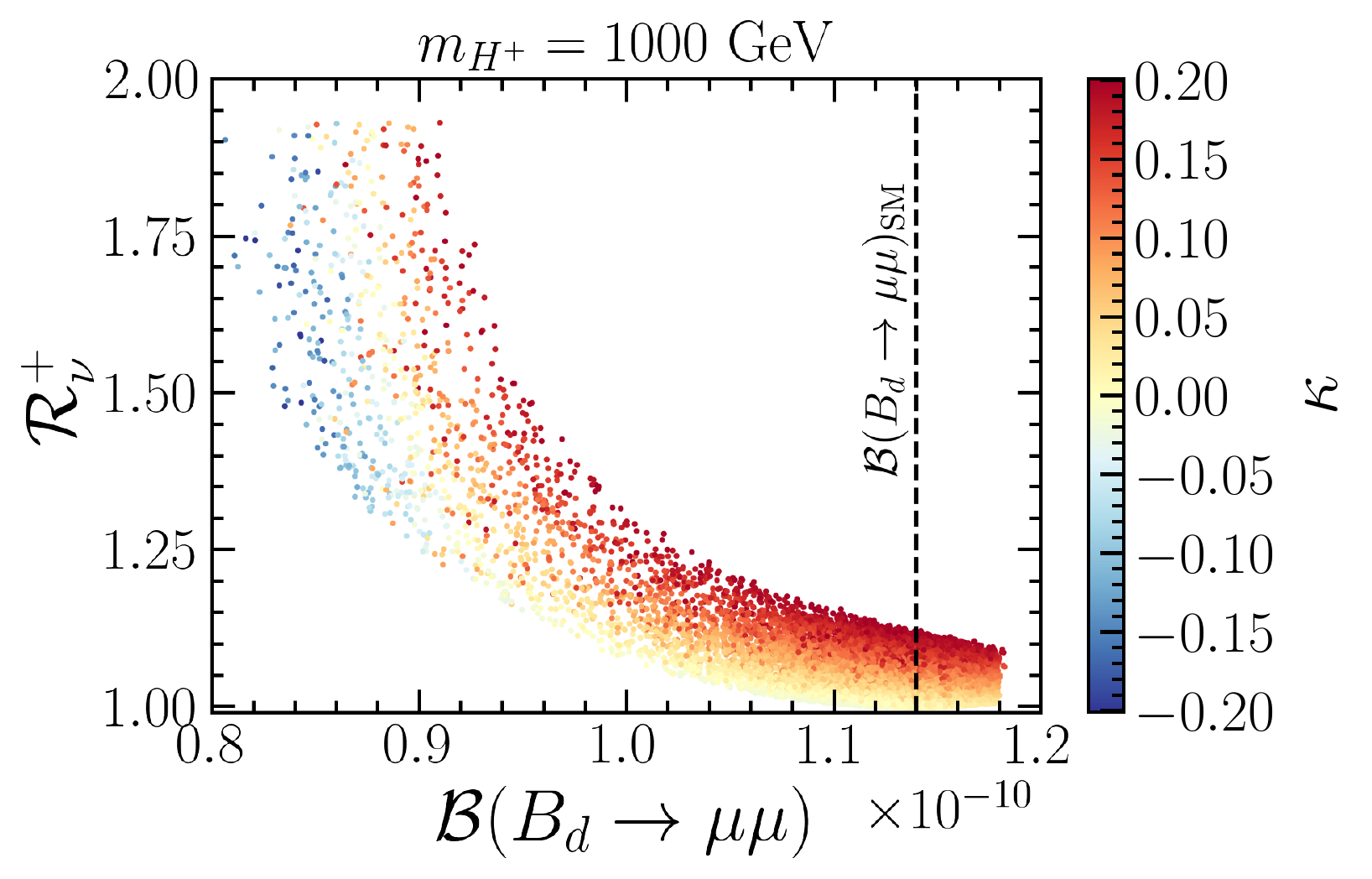}
\includegraphics[width=6.7cm, height=5.5cm]{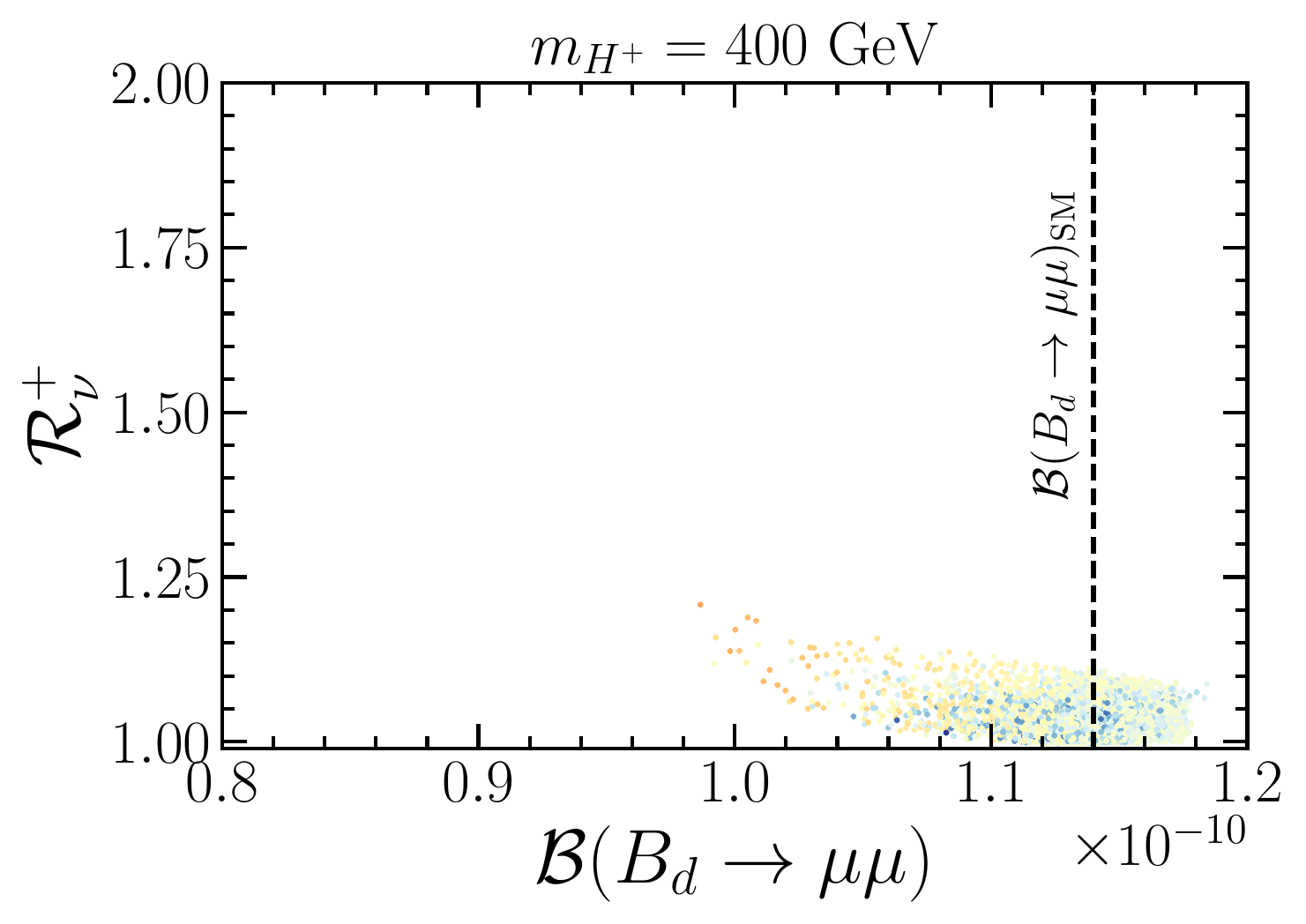}\includegraphics[width=8.4cm, height=5.5cm]{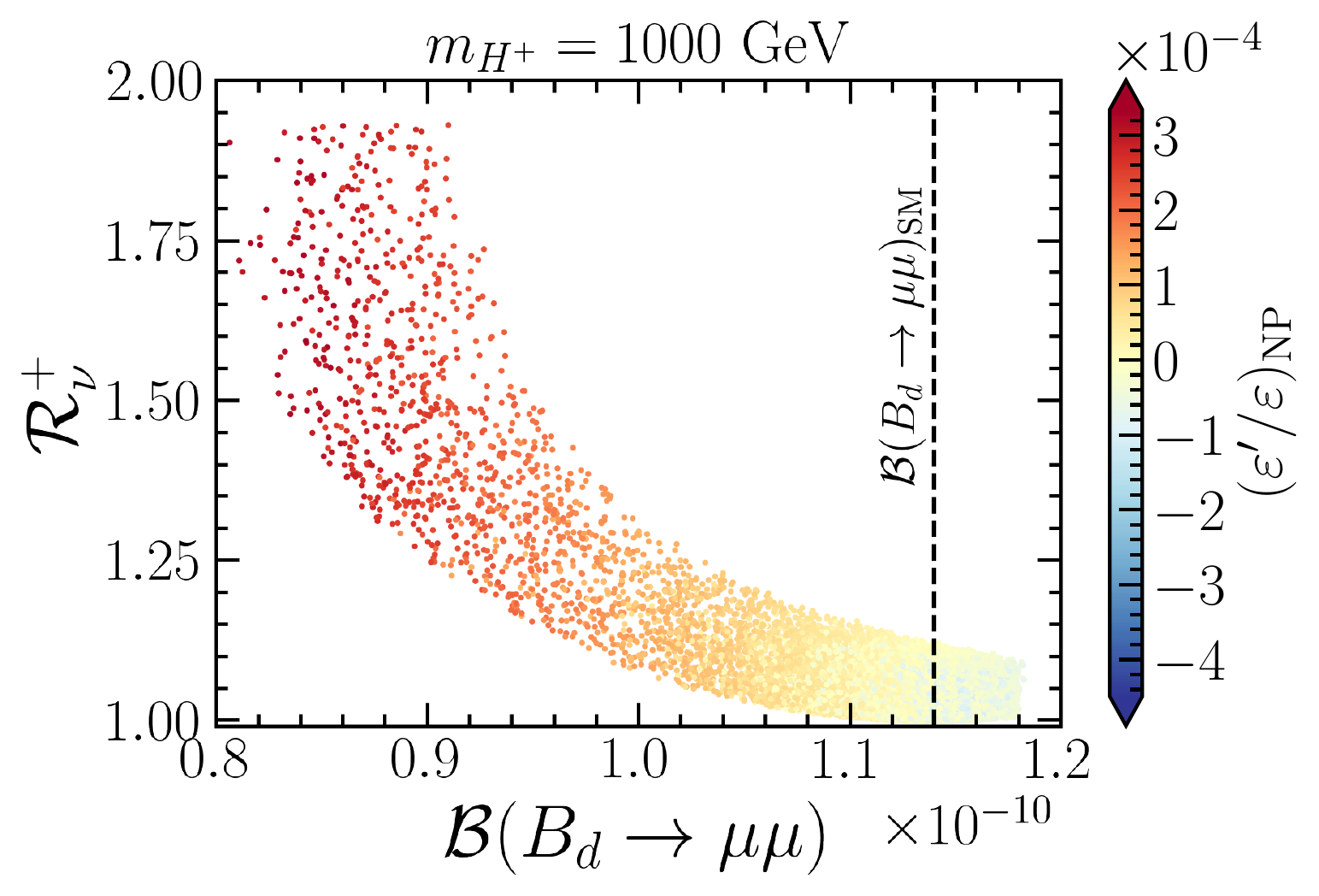}
\caption{Correlations between $\Kpnn$, $B_d\to\mu^+\mu^-$,
and $\kappa$ (upper) and $(\epsp)_{\rm NP}$ (lower).}
\label{fig: kpnn-Bdmm}
\end{figure}

Similar correlation is also observed for case of $B_d\to\mu\mu$,
as can be seen from Fig.~\ref{fig: kpnn-Bdmm}.
The decay is not measured yet; the latest limit from LHCb
\cite{LHCb:2021vsc, LHCb:2021awg} based on Full Run 1 and Run 2 data reads
${\cal B}(B_d\to\mu\mu) < 2.6 \times 10^{-10}$ at $95\%$ C.L,
while the recent analysis from CMS collaboration giving a more precise limit 
${\cal B}(B_d\to\mu\mu) < 1.9 \times 10^{-10}$ at $95\%$ C.L. \cite{CMS: ICHEP2022}.
For the corresponding SM prediction, we find ${\cal B}(B_d 
\to \mu\mu)_{\rm SM} = (1.14 \pm 0.12)\times 10^{-10}$, 
the central value of which is shown as black dashed line 
in Fig.~\ref{fig: kpnn-Bdmm}. 




\section{Discussion and Summary}\label{sec: summary}

We return to discuss briefly the implications of the 
new CMS result~\cite{CMS: ICHEP2022} of ${\cal B}(B_s 
\to \mu\mu)$. Our results for light $H^+$ remains 
unaffected, as this scenario was already tightly 
constrained by B sector observables and $\eps$. However, 
for heavy $H^+$, there are important changes. After 
imposing $2\sigma$ range of CMS value in our parameter 
scan, the most significant consequences are for $\eps$ 
and $\Kpnn$. For the former, we find $\kappa <0$ gets 
mostly ruled out, and limited now to $[-0.05, 0.2]$. For 
the latter, we find ${\cal R}_\nu^+$ can be enhanced 
only up to $50\%$, although we do find a few points 
reaching 80\%. For $\KLnn$, it can at best be suppressed 
by $15\%$ compared to SM. {No appreciable change for 
$(\epsp)_{\rm NP}$ is observed, and all correlations  discussed remain intact.}

{In this paper, we have studied NP effects of top-related 
extra Yukawa couplings, in particular highlighting the 
remarkable sensitivity of rare kaon decays in probing the FCNH 
coupling $\rho_{ct}$. In principle, if one allows for flavor 
violation in the down sector, significant NP effects are 
expected in $K^+\to \pi^+\nu\bar\nu$. This has already been 
pointed out previously, for example in 
Refs.~\cite{Isidori:2002qe,Isidori:2006jh} in context of 
supersymmetric (SUSY) models at large $\tan\beta$. In g2HDM, 
$H^+$ interaction terms concerning $\rho^d$ are given in  
Eq.~\eqref{eq: Lag}, 
and by the same top-$H^+$ loop diagrams as we presented, they 
will contribute to kaon observables. For example, for $K^+\to 
\pi^+\nu\bar\nu$, the $\rho^d$ couplings generate the effective 
operator $(\bar s\gamma_\mu R d)(\bar \nu \gamma_\mu 
\gamma_5\nu)$, with corresponding WC obtained from 
Eq.~\eqref{eq: b2snunu-WC} by substituting $(V^\dagger 
\r^u)_{2i}(\r^{u\dagger} V)_{i1} \to  - (\r^{d\dagger} 
\,V^\dagger)_{2i}(V \r^{d} )_{i1}$. Then flavor violating 
down couplings $\rho_{bq}$ $(q=s, d)$ will give leading 
contribution to the $s\to d \nu\bar\nu$ amplitude, since the 
concerned effect is proportional to $|V_{tb}|^2$ and hence 
not CKM-suppressed. However, these couplings induce $B_q$ 
mixing at {\it tree-level} via $H, A$ exchange, and therefore 
are constrained to be very small. Taking $m_H = m_A = 1$ TeV 
and $c_\gamma\to 0$, we find that $B_q$ mixing data (see 
Table~\ref{tab: B-meson-data}) gives the following $95\%$ C.L. 
bounds: $-1.7 < \rho_{bs}\rho_{sb}/10^{6} < 3.8$, $-0.2 < 
\rho_{bd}\rho_{db}/10^{6} < 1.4 $. If one assumes $\rho_{bq}$ 
are of similar strength as $\rho_{qb}$, then these bounds 
suggest that down-type flavor violating couplings are 
extremely tiny, at ${\cal O}(10^{-3})$. For lighter values of 
$m_H$, $m_A$, these bounds become only more stringent. With 
$\rho_{bq}$ coupling strengths as above, we find NP effects 
in $K \to \pi\nu\bar\nu$ from down-type flavor violation to be 
completely negligible ($< 1\%$).} 

Before offering our summary, we comment on FCNH $t \to 
ch$ decay, where CMS recently set the most stringent 
limit~\cite{CMS:2021gfa} of ${\cal B}(t \to ch) < 
0.094\%$ at $95\%$ C.L., based on $137~{\rm fb}^{-1}$ 
data at 13 TeV. This would put a constraint on the 
combination of $|c_\gamma\, \tilde \rho_{tc}|$, where 
$\tilde \rho_{tc} \equiv \sqrt{\rho_{tc}^2 + 
\rho_{ct}^2}/\sqrt{2}$. We find $|c_\gamma \,\tilde 
\rho_{tc}| < 0.059$, which can be evaded by having 
$c_\gamma$ small enough. 
One may think the ACME bound on electron EDM, $|d_e|$,
as the most challenging. But as alluded to in the
Introduction, a hierarchy between g2HDM diagonal Yukawa 
couplings of top and electron $|\r_{ee}/\rtt| \propto 
\l_e/\l_t$, which echoes the one seen already in SM 
Yukawa couplings helps one to handily 
evade~\cite{Fuyuto:2019svr} the ACME bound, by a 
couple orders of magnitude, which should be watched.

In summary, we explore g2HDM contributions of extra 
top Yukawa couplings $\rho_{ij}$ to several kaon 
processes, including kaon mixing, $\epsp$, and rare 
$\Kpnn$, $\KLnn$, and $K_{L, S}\to \mu\mu$ decays. 
We first point out that $\eps$ provides significant 
constraint on $\rho_{ij}$ couplings that are 
complementary to the B sector, giving the leading
constraint on the off-diagonal $\rct$ coupling. We 
consider two disparate masses of $H^+$: 400 GeV and 
1000 GeV. We find g2HDM contribution to $(\epsp)_{\rm 
NP}$ as large as $\sim (1$--$3) \times 10^{-4}$ are 
achievable while satisfying B sector and $\eps$ 
constraints, with light $H^+$ preferring negative 
values reaching down to $\sim -5 \times 10^{-4}$. For 
rare $K \to \pi \nu\nu$ decays, we find opposing 
effects in charged vs neutral modes. For light $H^+$ 
case, we find $\Kpnn$ can be enhanced by up to $\sim 
20\%$, while $\KLnn$ can receive suppression up to 
$\sim 10\%$. However, for $m_{H^+} = 1000$ GeV, the 
B physics and $\eps$ constraints on $\rho_{ij}$ become 
weaker, and sizable enhancements of kaon decays become 
possible. We find that $\Kpnn$ can easily saturate the 
current NA62 bound, while $\KLnn$ can be suppressed by 
up to $20\%$ over the SM. For $K_L\to\mu\mu$, large 
theory errors make it ineffective as a probe for $H^+$ 
effects, while $\KSmm$ remains SM-like in g2HDM. 
Exploring the correlation of $\Kpnn$ with $B_s \to 
\mu^+\mu^-$, we find for heavy $m_{H^+} = 1000$ GeV, 
enhanced $\Kpnn$ in g2HDM implies suppression of 
$B_s \to \mu\mu$. Precise measurements of ${\cal 
B}(\Kpnn)$ and ${\cal B}(B_s \to \mu^+\mu^-)$ should 
be able to distinguish the parameter space 
corresponding to sub-TeV vs TeV scale $H^+$.

\vskip0.2cm
\noindent{\bf Acknowledgments} 
{This research is supported by 
MOST 
110-2639-M-002-002-ASP 
of Taiwan, and NTU grants 110L104019 and 110L892101.}
 
\appendix
\section{Loop Functions}{\label{app: loop}}
The loop functions related to $|\Delta F|=2$ processes are
\cite{Crivellin:2019dun},
\begin{align}
  F_1(a, b) &= \frac{-1}{(1-a)(1-b)}
  + \frac{b^2 \log b}{(1-b)^2(a-b)}
  - \frac{a^2 \log a}{(1-a)^2 (a-b)},\\
  F_2(a, b, c) &= \frac{-3 a^2 \log a}{(a-1)(a-b)(a-c)}
  + \frac{b(4a - b)\log b}{(b-1)(a-b)(b-c)}
  + \frac{c(4a -c) \log c}{(c-1)(a-c)(c-b)}.
\end{align}
The loop functions related to $|\Delta F|=1$ processes
$s \to d f\bar f$ ($f = q, \ell, \nu$) are \cite{Iguro:2017ysu,Iguro:2019zlc},
\begin{align}
  G_{\g 1}(a) &= -\frac{16-45 a + 36 a^2 - 7 a^3 + 6(2-3a)\log a}
  {36 (1-a)^4},\\
  G_{\g 12}(a) &= -\frac{2 -9a + 18 a^2 - 11 a^3 + 6a^3\log a}
  {36 (1-a)^4} + \frac{2}{3} G_{\g 1}(a),\\
  G_Z(a) &= \frac{a(1-a+ \log a)}{2 (1-a)^2},
  \end{align}
and function related to $s \to d g$ is \cite{Iguro:2017ysu},
\begin{align}
    F_{\sigma}(a) = -\frac{2+3a-6a^2+a^3+6a\log a}{12(1-a)^4}.
\end{align}

\section{\boldmath Neutral {\it B} Meson 
Mixings and Mixing-induced CPV}
{\label{app: acp}}

The $\Delta B=2$ effective Hamiltonian relevant for our purpose is given by,
\begin{equation}
  {\cal H}_{\rm eff}(\Delta B=2)
  = (C_{HH}^{(q)}+C_{WH}^{(q)})\, (\bar{b} \gamma^{\mu} P_L q)(\bar{b} \gamma^{\mu} P_L q) + {\rm H.c.},
  \label{eq: Heff-Bmixing}
\end{equation}
where $q=s, d$ corresponds to $B_s$- and $B_d$-mixing, respectively.
The coefficients $C_{HH}^{(q)}$ and $C_{WH}^{(q)}$ are same as given in
Eqs.~\eqref{eq: CHH-K} and \eqref{eq: CWH-K} after obvious change
of flavor indices.

The matrix element for $\bar B_q$--$B_q$ mixing is defined as
$M^{q\ast}_{12} = \langle \bar B_q | {\cal H}_{\rm eff}(\Delta B=2)|B_q \rangle$, where $M_{12}$ is a complex quantity:
$M_{12}\equiv |M_{12}| e^{2 i \phi _q}$.
Then absolute value of $M_{12}$ determines the neutral $B_q$ mass difference as,
\begin{align}
  \Delta M_{B_q} = 2 |M^{q}_{12}| \equiv  2 |M^{q}_{12}(\rm SM) + M^{q}_{12}(\rm NP)|,
\end{align}
while phases $\phi_q$ are convention dependent quantities and
defined following Ref.~\cite{Buras:2013ooa} as,
\begin{align}
  \phi_s =  \beta_s + \phi_s^{\rm NP} \quad \phi_d =  \beta_d + \phi_d^{\rm NP},
\end{align}
with $\beta_s$ and $\beta$ given by CKM elements, $V_{ts} = - |V_{ts}| e^{-i \beta_s}$
and $V_{td} =  |V_{td}| e^{-i \beta}$, and $\phi_q^{\rm NP}$ is due to contribution from
NP effective Hamiltonian in Eq.~\eqref{eq: Heff-Bmixing}.
{The SM predictions~\cite{DiLuzio:2019jyq} for mass differences are
$\Delta M_{B_s} = (18.4^{+0.7}_{-1.2})~{\rm ps}^{-1}$
and $\Delta M_{B_d} = (0.533^{+0.022}_{-0.036})~{\rm ps}^{-1}$,
and the corresponding experimental values are given in Table~\ref{tab: B-meson-data}.}

The mixing phases $\phi_q$ are inferred from the measurement of mixing-induced CP asymmetries $S_{\psi  K_S}$ and $S_{\psi \phi}$ which appear as coefficients in the
time-dependent asymmetries of
$B_d \to\psi K_S$ and $B_s \to \psi\phi $,
\begin{align}
    A^{\psi K_S}_{\rm CP}
    = S_{\psi K_S} \sin (\Delta M_d t),
    \quad
    A^{\psi \phi}_{\rm CP}
    = S_{\psi \phi} \sin (\Delta M_s t).
\end{align}
where in presence of NP phase $\phi_q^{\rm NP}$ the CP asymmetries
$S_{\psi  K_S}$ and $S_{\psi \psi}$ are  given by \cite{Buras:2013ooa},
\begin{align}
  S_{\psi K_S} = \sin(2 \beta + 2 \phi^{\rm NP}_d),
  \quad
  S_{\psi \phi} = \sin(2 |\beta_s| - 2 \phi^{\rm NP}_s).
\end{align}

\end{document}